\documentclass[twocolumn, twocolappendix]{aastex701}

\newcommand\zphot{$z_{\text{phot}}$}

\begin{document}

\title{JWST Advanced Deep Extragalactic Survey (JADES) Data Release 5: Catalogs of inferred morphological properties of galaxies from JWST/NIRCam imaging in GOODS-N and GOODS-S}


\author[0000-0001-6301-3667]{Courtney Carreira}
\affiliation{Department of Astronomy and Astrophysics, University of California, Santa Cruz, 1156 High Street, Santa Cruz, CA 95064, USA}
\email{ccarreir@ucsc.edu} 

\author[orcid=0000-0002-4271-0364]{Brant E. Robertson} 
\affiliation{Department of Astronomy and Astrophysics, University of California, Santa Cruz, 1156 High Street, Santa Cruz, CA 95064, USA}
\email{brant@ucsc.edu}

\author[0000-0002-9708-9958]{A.\ Lola Danhaive}
\affiliation{Kavli Institute for Cosmology, University of Cambridge, Madingley Road, Cambridge, CB3 0HA, UK}
\affiliation{Cavendish Laboratory - Astrophysics Group, University of Cambridge, 19 JJ Thomson Avenue, Cambridge, CB3 0HE, UK}
\email{ald66@cam.ac.uk}

\author[0000-0001-7673-2257]{Zhiyuan Ji}
\affiliation{Steward Observatory, University of Arizona, 933 N. Cherry Avenue, Tucson, AZ 85721, USA}
\email{zhiyuanji@arizona.edu}

\author[0000-0002-7893-6170]{Marcia Rieke}
\affiliation{Steward Observatory, University of Arizona, 933 N. Cherry Avenue, Tucson, AZ 85721, USA}
\email{mrieke@gmail.com}

\author[0000-0002-8224-4505]{Sandro Tacchella}
\affiliation{Kavli Institute for Cosmology, University of Cambridge, Madingley Road, Cambridge, CB3 0HA, UK}
\affiliation{Cavendish Laboratory - Astrophysics Group, University of Cambridge, 19 JJ Thomson Avenue, Cambridge, CB3 0HE, UK}
\email{st578@cam.ac.uk}

\author[0000-0001-6917-4656]{Natalia C.\ Villanueva}
\affiliation{Department of Astronomy, The University of Texas at Austin, Austin, TX 78712, USA}
\email{nataliavillanueva@utexas.edu}

\author[0000-0001-9262-9997]{Christopher N.\ A.\ Willmer}
\affiliation{Steward Observatory, University of Arizona, 933 N. Cherry Avenue, Tucson, AZ 85721, USA}
\email{cnaw@as.arizona.edu}

\author[0000-0002-8876-5248]{Zihao Wu}
\affiliation{Center for Astrophysics $|$ Harvard \& Smithsonian, 60 Garden Street, Cambridge, MA 02138, USA}
\email{zihao.wu@cfa.harvard.edu}

\author[0000-0003-3307-7525]{Yongda Zhu}
\affiliation{Steward Observatory, University of Arizona, 933 N. Cherry Avenue, Tucson, AZ 85721, USA}
\email{yongdaz@arizona.edu}

\author[0000-0003-0215-1104]{William M.\ Baker}
\affiliation{DARK, Niels Bohr Institute, University of Copenhagen, Jagtvej 155A, DK-2200 Copenhagen, Denmark}
\email{william.baker@nbi.ku.dk}

\author[0000-0002-8651-9879]{Andrew J.\ Bunker}
\affiliation{Department of Physics, University of Oxford, Denys Wilkinson Building, Keble Road, Oxford OX1 3RH, UK}
\email{andy.bunker@physics.ox.ac.uk}

\author[0000-0002-0450-7306]{Alex J.\ Cameron}
\affiliation{Cosmic Dawn Center (DAWN), Copenhagen, Denmark}
\affiliation{Niels Bohr Institute, University of Copenhagen, Jagtvej 128, DK-2200, Copenhagen, Denmark}
\email{alex.cameron@nbi.ku.dk}

\author[0000-0002-7636-0534]{Jacopo Chevallard}
\affiliation{Department of Physics, University of Oxford, Denys Wilkinson Building, Keble Road, Oxford OX1 3RH, UK}
\email{chevalla@iap.fr}

\author[0000-0002-9551-0534]{Emma Curtis-Lake}
\affiliation{Centre for Astrophysics Research, Department of Physics, Astronomy and Mathematics, University of Hertfordshire, Hatfield AL10 9AB, UK}
\email{e.curtis-lake@herts.ac.uk}

\author[0009-0009-8105-4564]{Qiao Duan}
\affiliation{Kavli Institute for Cosmology, University of Cambridge, Madingley Road, Cambridge, CB3 0HA, UK}
\affiliation{Cavendish Laboratory - Astrophysics Group, University of Cambridge, 19 JJ Thomson Avenue, Cambridge, CB3 0HE, UK}
\email{qd231@cam.ac.uk}

\author[0000-0002-2929-3121]{Daniel J.\ Eisenstein}
\affiliation{Center for Astrophysics $|$ Harvard \& Smithsonian, 60 Garden Street, Cambridge, MA 02138, USA}
\email{deisenstein@cfa.harvard.edu}

\author[0000-0003-4565-8239]{Kevin Hainline}
\affiliation{Steward Observatory, University of Arizona, 933 N. Cherry Avenue, Tucson, AZ 85721, USA}
\email{kevinhainline@arizona.edu}

\author[0000-0002-8543-761X]{Ryan Hausen}
\affiliation{Department of Physics and Astronomy, The Johns Hopkins University, 3400 N. Charles Street, Baltimore, MD 21218, USA}
\email{rhausen@ucsc.edu}

\author[0000-0002-9280-7594]{Benjamin D.\ Johnson}
\affiliation{Center for Astrophysics $|$ Harvard \& Smithsonian, 60 Garden Street, Cambridge, MA 02138, USA}
\email{benjamin.johnson@cfa.harvard.edu}

\author[0000-0002-4985-3819]{Roberto Maiolino}
\affiliation{Kavli Institute for Cosmology, University of Cambridge, Madingley Road, Cambridge, CB3 0HA, UK}
\affiliation{Cavendish Laboratory - Astrophysics Group, University of Cambridge, 19 JJ Thomson Avenue, Cambridge, CB3 0HE, UK}
\affiliation{Department of Physics and Astronomy, University College London, Gower Street, London WC1E 6BT, UK}
\email{rm665@cam.ac.uk}

\author[0009-0006-4365-2246]{Petra Mengistu}
\affiliation{Department of Astronomy and Astrophysics, University of California, Santa Cruz, 1156 High Street, Santa Cruz, CA 95064, USA}
\email{pmengist@ucsc.edu}

\author[0000-0001-8630-2031]{Dávid Puskás}
\affiliation{Kavli Institute for Cosmology, University of Cambridge, Madingley Road, Cambridge, CB3 0HA, UK}
\affiliation{Cavendish Laboratory - Astrophysics Group, University of Cambridge, 19 JJ Thomson Avenue, Cambridge, CB3 0HE, UK}
\email{dp670@cam.ac.uk}

\author[0000-0002-5104-8245]{Pierluigi Rinaldi}
\affiliation{Space Telescope Science Institute, 3700 San Martin Drive, Baltimore, MD 21218, USA}
\email{prinaldi@stsci.edu}

\author[0000-0001-6561-9443]{Yang Sun}
\affiliation{Steward Observatory, University of Arizona, 933 N. Cherry Avenue, Tucson, AZ 85721, USA}
\email{sunyang@arizona.edu}

\author[0000-0002-9081-2111]{James A.\ A.\ Trussler}
\affiliation{Center for Astrophysics $|$ Harvard \& Smithsonian, 60 Garden Street, Cambridge, MA 02138, USA}
\email{james.trussler@cfa.harvard.edu}

\author[0000-0003-4891-0794]{Hannah \"Ubler}
\affiliation{Max-Planck-Institut f\"ur extraterrestrische Physik (MPE), Gie{\ss}enbachstra{\ss}e 1, 85748 Garching, Germany}
\email{hannah@mpe.mpg.de}

\author[0000-0001-7658-9710]{Anavi Uppal}
\affiliation{Department of Astronomy and Astrophysics, University of California, Santa Cruz, 1156 High Street, Santa Cruz, CA 95064, USA}
\email{anuppal@ucsc.edu}

\author[0000-0003-2919-7495]{Christina C.\ Williams}
\affiliation{NSF National Optical-Infrared Astronomy Research Laboratory, 950 N. Cherry Avenue, Tucson, AZ 85719, USA}
\email{christina.williams@noirlab.edu}

\correspondingauthor{Courtney Carreira}
\email[show]{ccarreir@ucsc.edu}


\begin{abstract}

We present morphological parameters and their uncertainties for all sources detected in JWST/NIRCam imaging in GOODS-N and GOODS-S from the JWST Advanced Deep Extragalactic Survey (JADES) catalogs. We model the surface brightness profiles of these sources with single-component Sérsic profiles, performing Bayesian inference of galaxy structural parameters. We fit each of the $>10^5$ sources with every available JWST/NIRCam wide-band filter individually, amounting to over 3 million S\'ersic profiles computed. We provide catalogs of this morphological information, building one of the largest extragalactic morphological datasets to date, which we share alongside imaging and photometry from the JADES Data Release 5. With this information, we analyze the rest-frame optical redshift evolution of the effective radius and the surface luminosity density within a radius of 1 kiloparsec, $\Sigma_{\text{1 kpc}}$, for 24,692 galaxies at $z>1$. We find $r_{\text{eff}} \propto (1+z)^{-0.635 \pm 0.013}$ kpc, while $\Sigma_{\text{1 kpc}}$ is relatively constant across time. Additionally, we explore bulge-disk decomposition on a subset of 8,390 galaxies in the JADES deep imaging covering the Hubble Ultra Deep Field, finding the effective radius of the bulge-components to increase marginally with time, whereas the disk-component sizes evolve as $r_{\text{eff,disk}} \propto (1+z)^{-1.091 \pm 0.043}$. Future work modeling multi-component surface brightness profiles will enable further analysis of the morphological evolution of galaxies across cosmic time.

\end{abstract}


\section{Introduction} \label{sec:intro}

Galaxy morphology is a key tracer of the evolutionary history of galaxies across cosmic time. The physical processes which govern the growth and disruption of galaxies are imprinted on their structure; therefore, tracing the evolution of the morphology of galaxies can shed light on which processes govern galaxy growth at different epochs (for a comprehensive review, see \citealt{conselice_2014}).

In recent decades, large extragalactic surveys have expanded the volume of galaxies whose morphologies we have characterized and connected to other galaxy properties. On the ground, the Sloan Digital Sky Survey (SDSS; \citealt{sdss_overview}) observed a statistical census of galaxies in the local Universe, enabling a broad investigation of the effects of luminosity, morphology, and stellar mass on the size distribution of galaxies. \citet{shen_sloan_2003} found the size distribution at fixed luminosity is well-described by a log-normal function and, having found different scaling relations between early- versus late-type galaxies, suggested that early- and late-type galaxies undergo different formation and assembly histories. Complementarily to SDSS, the \textit{Hubble Space Telescope} (HST) conducted several extragalactic surveys which imaged galaxies with high-resolution in the rest-frame optical and ultraviolet (UV). These surveys, such as the Cosmic Assembly Near-infrared Deep Extragalactic Legacy Survey (CANDELS; \citealt{grogin_candels_2011, koekemoer_candels_2011}) and Cosmic Evolution Survey (COSMOS; \citealt{cosmos_overview_2007, cosmos_hst_2007}) dramatically altered our understanding of the evolution of intermediate-redshift ($z \sim 3$) galaxies. These works established that intermediate-redshift galaxies are typically smaller than their lower-redshift counterparts of equal mass (e.g., \citealt{vdw_2014}). Furthermore, early work with HST demonstrated that while most $1 \lesssim z \lesssim 3$ galaxies are morphologically irregular (e.g., \citealt{conselice_2011}), some sufficiently massive galaxies can settle into Hubble types in the early universe \citep{bruce_2012, mortlock_2013}. With these insights, it became clear that classifying high-redshift galaxies requires a combination of physical and morphological identifiers to fully explain the diverse population of galaxies we can now observe. Over time, these surveys collectively formed a statistical sample of high-redshift galaxies, spanning $z=0-10$, whose size evolution was characterized by \citet{shibuya_2015}.

With HST, observational analysis of high-redshift galaxies in optical light was limited to $z \lesssim 3$, beyond which only the rest-frame UV can be probed. UV observations primarily trace active star formation and young stellar populations, and are highly susceptible to dust attenuation. Therefore, to characterize older stellar populations and more broadly understand the properties of galaxies beyond $z \sim 3$, we require near-infrared observations.

The \textit{James Webb Space Telescope} (JWST) has revolutionized our ability to study galaxy evolution at high redshifts with deep, high-resolution near-infrared imaging of $z>3$ galaxies, tracing the rest-frame optical out to $z \sim 7$. Several studies have examined morphological trends  at $z>5$ in large surveys using JWST (e.g., \citealt{kartaltepe_2023, ormerod_2024, varadaraj_2024, martorano_2024, ward_2024, morishita_2024, huertas-company_2024, miller_2025, yang_2025, allen_2025, martorano_2025, genin_2025, danhaive_2025}). With these surveys, we have identified high-redshift galaxies with established disk and spheroid morphologies, suggesting that such features can form very early \citep{baker_2025}. Nonetheless, the population of clumpy, irregular morphologies remains significant at $z > 3$ \citep{hainline_2024, huertas-company_2024, delavega_2025, Zhu2026JADESClumps}.

The JWST Advanced Deep Extragalactic Survey (JADES; PIs: M. Rieke and N. L\"utzgendorf; \citealt{bunker_jades_2020, jades_overview, rieke_2023}) is a collaboration of the Near-Infrared Camera (NIRCam; \citealt{rieke_nc, rieke_nc_recent}) and Near-Infrared Spectrograph (NIRSpec; \citealt{jakobsen_ns}) Science Development Teams, conducting deep imaging and spectroscopy in two legacy fields, Great Observatories Origins Deep Survey North and South (GOODS-N and GOODS-S, respectively; \citealt{goods_2004}). JADES covers over 400 arcmin\textsuperscript{2} across both GOODS-N and GOODS-S, resulting in the detection of $>10^5$ sources spanning $0 < z< 15$.

The depth of imaging allows for resolved morphological analysis on faint galaxies, down to $\sim 30$ AB mag in some regions \citep{Johnson2026, Robertson2026}. To understand the interplay of galaxy morphology and evolution, the characteristics of the JADES imaging are particularly ideal; with JADES, we can conduct a uniform analysis of the morphology of galaxies that are non-uniform in their individual characteristics, representing diverse populations in magnitude, redshift, stellar mass, and star formation activity.

To systematically characterize the evolution of galaxy morphology over time, we require analytical models that can capture the features of each galaxy's surface brightness profile. \citet{sersic_1963} described the S\'ersic profile, a generalization of the earlier de Vaucouleurs profile \citep{deV_profile}, to model the surface brightness of a galaxy as $R^{1/n}$, with the S\'ersic index $n$ representing the concentration of the light profile. The S\'ersic profile has remained the dominant parametric model for representing the surface brightness profiles of galaxies, allowing for the uniform analysis of structural parameters in modern imaging of galaxies.
In this work, we provide catalogs of morphological properties for over 3 million S\'ersic profiles, fitting the individual surface brightness profiles of $>10^5$ sources in each of the eight JWST/NIRCam wide-band filters. To characterize the surface brightness profiles of these sources, we fit single-component S\'ersic profiles with \texttt{pysersic} \citep{pysersic}, a \textsc{Python} package which performs Bayesian inference to characterize the morphology of these galaxies. This method allows us to perform robust uncertainty estimation on the returned structural parameters, which include the half-light (effective) radius, S\'ersic index, and axis ratio. We perform this modeling separately for all available JWST/NIRCam wide-band filters; therefore, we provide size measurements in a variety of rest-frames, each probing different physical characteristics of the galaxy. The catalogs contain the posterior distributions for the structural parameters of the S\'ersic profile, with the goal of providing the community with the information required to perform additional statistical analyses on this dataset. To demonstrate the power of this information, we characterize the redshift evolution of these morphological parameters for the JADES survey.

This paper has the following structure. In \S\ref{sec:data}, we describe the JADES imaging mosaics, data reduction, and photometry. We detail the methodology for modeling the surface brightness profiles of these galaxies with \texttt{pysersic} in \S\ref{sec:modeling}. In \S\ref{sec:catalog}, we describe the JADES morphological catalogs presented here, including their content and format, and discuss caveats regarding the accuracy of the presented S\'ersic models. With these catalogs, in \S\ref{sec:redshift_evolution}, we present our results on the size--redshift evolution at $z>1$, as well as the redshift evolution of bulge-disk decomposition profiles for a subset of galaxies; we discuss these results in \S\ref{sec:discussion}. Lastly, in \S\ref{sec:conclusions}, we summarize our work. 

Throughout this paper, we adopt a \citet{planck} $\Lambda$CDM cosmology with $H_0 = 67.4$ km s\textsuperscript{-1} Mpc\textsuperscript{-1} and $\Omega_{m,0} = 0.315$. All magnitudes are expressed in the AB system \citep{oke_1974, abmags}.

\section{Data} \label{sec:data}

The JADES imaging survey combines approximately 800 hours of JWST/NIRCam guaranteed time observations across GOODS-N and GOODS-S. In addition to JADES programs, we include other JWST/NIRCam imaging in these fields to produce the imaging mosaics described in \citet{Johnson2026} and utilized in this work. In GOODS-S, there is imaging coverage in all JWST/NIRCam wide-band filters (F070W, F090W, F115W, F150W, F200W, F277W, F356W, and F444W), as well as coverage in all but two medium-band filters (F162M, F182M, F210M, F250M, F300M, F335M, F410M, F430M, F460M, and F480M). The GOODS-N footprint also includes all of the JWST/NIRCam wide-band filters, but with less area covered overall. For the medium-band filters in GOODS-N, there are two fewer filters than in GOODS-S, including only F162M, F182M, F210M, F300M, F335M, F410M, F430M, and F460M. In this paper, we briefly describe the observations and imaging reduction, photometric catalogs, photometric redshift estimation, and model point spread functions (mPSFs) from JADES Data Release 5 (DR5); these methods are detailed in the JADES DR5 companion papers by \citet{Johnson2026} and \citet{Robertson2026}.

\subsection{JWST/NIRCam Imaging}\label{subsec:imaging}

In this work, we analyze the JADES DR5 imaging mosaics presented in \citet{Johnson2026}. Here, we provide a brief description of these mosaics.

The JWST/NIRCam imaging mosaics primarily utilize JADES observational programs; these include JWST Programs 1180, 1181, 1210, 1286, 1287, and 4540 \citep{jades_overview}, and the JADES Origins Field (JOF) Program 3215 \citep{jof}.

We also include other surveys which have observed GOODS-N and GOODS-S. These additional surveys include the Next Generation Deep Extragalactic Exploratory Public (NGDEEP) Survey \citep{ngdeep}, the JWST Extragalactic Medium-band Survey (JEMS; \citealt{jems}), the Prime Extragalactic Areas for Reionization and Lensing Science (PEARLS) Survey \citep{pearls}, the First Reionization Epoch Spectroscopically Complete Observations (FRESCO) Survey \citep{fresco}, the Parallel wide-Area Nircam Observations to Reveal And Measure the Invisible Cosmos (PANORAMIC) Survey \citep{panoramic}, the Complete NIRCam Grism Redshift Survey (CONGRESS; \citealt{congress}), the Observing All phases of StochastIc Star formation (OASIS) survey \citep{oasis}, the MIRI Deep Imaging Survey (MIDIS; \citealt{midis}), the Bias-free Extragalactic Analysis for Cosmic Origins with NIRCam (BEACON) survey \citep{beacon}, the Slitless Areal Pure-Parallel High-Redshift Emission Survey (SAPPHIRES, \citealt{sapphires}), and lastly, the Public Observation Pure Parallel Infrared Emission-Line Survey (POPPIES; \citealt{poppies}). Other public JWST imaging in GOODS-N and GOODS-S is also included; a full list and descriptions of these programs can be found in \citet{Johnson2026}.

For a complete description of the imaging reduction, we refer to \citet{Johnson2026}; a brief overview is provided here. The JADES imaging reduction pipeline relies on the \texttt{jwst} pipeline to fit the raw ramp data, produce astrometrically-calibrated and background-subtracted images from the count rate images, resample, and combine the images into mosaics. Several custom reduction steps are also utilized; these steps include a correction for crosstalk correlation, 1/$f$ noise fitting
and removal, wisp and persistence template fitting and removal, custom long-wavelength sky-flats generated from observed data, and a mosaic outlier rejection step intended to preserve the cores of compact, unsaturated sources.

In this work, we utilize the JADES DR5 imaging mosaics which are at their native resolution and have not been convolved to a single instrumental resolution. The pixel scale of these mosaics is 0.03$\arcsec$ pixel\textsuperscript{-1}, as described in \citet{Johnson2026}.

\subsection{Photometric catalogs}\label{subsec:photometry}

The source detection and photometry performed on these imaging mosaics are described in \citet{Robertson2026}. In brief, sources are identified using a detection image, generated as inverse-variance-weighted signal-to-noise ratio (SNR) stacks of the long-wavelength imaging mosaics presented in \citet{Johnson2026}. A series of customized \texttt{photutils} \citep{photutils} routines are then deployed to generate segmentation maps, defining the pixels associated with the detected sources. Deblending of additional sources is performed by algorithmically identifying locally-dominant peaks in the SNR mosaic captured within segmentations. Additional curation of the source detection is performed manually to address some failure modes, such as diffraction spike shredding.

The source catalog is generated using methods similar to those employed by \texttt{photutils}. Centroids are computed as the SNR-weighted barycenter surrounding a locally dominant peak. A Gaussian regression model is deployed to estimate source properties, analytically solving for the source size and Kron radius \citep{kron_1980}.

\subsection{Model point spread functions}\label{subsec:psf}

The construction of the mPSFs utilized in this work are described in \citet{Johnson2026}, derived from the methods introduced by \citet{ji_2024_mpsf}. As described in this section, the JADES imaging mosaics incorporate the imaging of several distinct surveys. This being the case, we construct mPSFs for each filter mosaic, acquired at a specific position angle, in the GOODS-N and GOODS-S fields. We begin by simulating PSFs for a given filter and position angle, utilizing \texttt{STPSF} \citep{perrin_2014}. These PSFs are arrayed across the sky as simulated stars, matching the dither pattern of the individual observational programs used to create the mosaics. Using the \texttt{EPSFBuilder} routines from \texttt{photutils}, the mPSF for each mosaic is measured from these simulated stars. These mPSFs are then circularly apodized, as described in \citet{Robertson2026}.

\subsection{Photometric redshift estimation with EAZY}\label{subsec:eazy}

The methodology for calculating photometric redshifts is described in \citet{Robertson2026}; we provide a brief summary here. We follow the methods described in \citet{rieke_2023} and \citet{hainline_2024}, utilizing \texttt{EAZY} \citep{brammer_2008} for template-based photometric redshift estimates. We use the template set from \citet{hainline_2024}, and do apply photometric offsets to the JWST/NIRCam photometry for fitting. Redshift estimates are made with both circular apertures ($r=0.1\arcsec$) and Kron convolved ellipsoidal apertures; in this work, we adopt the circular aperture redshift estimates. The best photometric redshift is recorded as $z_a$ based on the minimum of the $\chi^2$ surface curve.

\section{Surface Brightness Profile Modeling} \label{sec:modeling}

We perform surface brightness profile modeling on all sources detected in JWST/NIRCam imaging in both GOODS-N and GOODS-S, as reported in the JADES DR5 photometric catalog associated with each field by \citet{Robertson2026}.

\begin{figure*}[t]
    \centering
    \includegraphics[width=\textwidth]{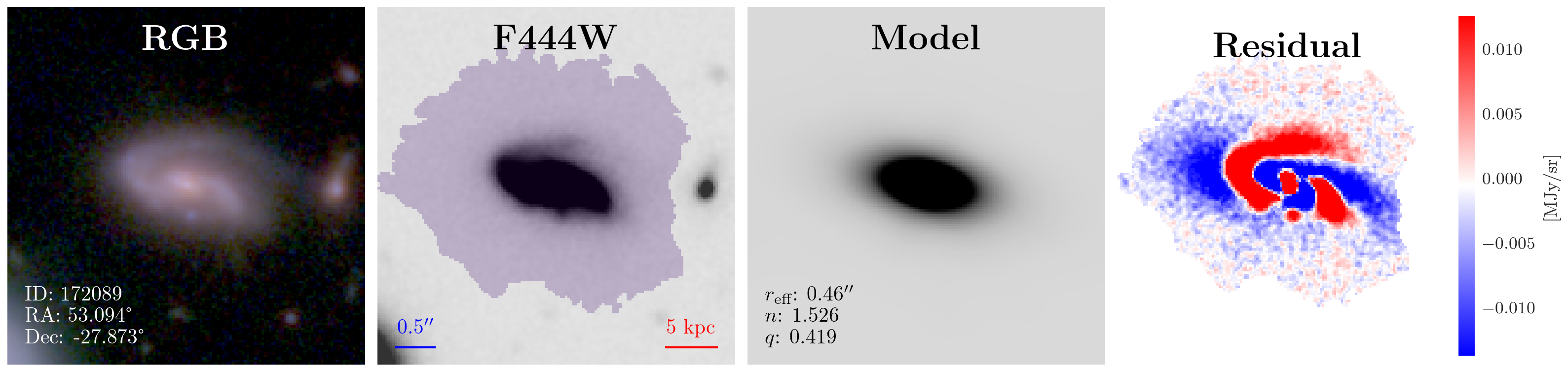}
    \caption{Example of a successful single-component S\'ersic fit to a spiral galaxy in GOODS-S. \textit{From left to right:} (1, left) F444W/F200W/F090W false-color RGB image of JADES ID 172089, with \zphot\ = 0.81 \citep{Robertson2026}. (2) F444W image of this object. Scale bars along the bottom indicate the size in angular and physical units, in blue and red, respectively. The purple shaded region represents the segmentation associated with this object. (3) The best-fit S\'ersic model for this object, as sampled by \texttt{pysersic}. The best-fit model is described the effective radius $r_{\text{eff}}$, S\'ersic index $n$, and axis ratio $q$ listed in the image. (4, right) The residual of subtracting the S\'ersic model from the F444W image. Underlying features, such as the spiral structure of the galaxy and a bright knot directly below the galactic nucleus, become clearly visible in the residual image.}
    \label{fig:example}
\end{figure*}

To perform this modeling, we use \texttt{pysersic} \citep{pysersic}, a Bayesian fitting framework which performs inference of galaxy structural parameters by using \texttt{JAX} \citep{jax2018github} and \texttt{NumPyro} \citep{numpyro_2021, numpyro_2025} for efficient and robust posterior distribution estimation. With \texttt{pysersic}, we fit S\'ersic profiles \citep{sersic_1963,sersic_1968} to the sources, described by
\begin{equation}
    I(R) = I_e \, \exp \left\{ -b_n \left[ \left( \frac{R}{R_e} \right) ^{1/n} -1 \right] \right\},
\end{equation}
where $I(R)$ is the intensity at a radius $R$, $I_e$ is the intensity at the half-light radius $R_e$, $n$ is the S\'ersic index, and $b_n$ is an analytical function of $n$ \citep{bn_approx}. 

\subsection{Producing cutouts of sources}\label{subsec:cutouts}

For each source, in any given JWST/NIRCam filter, we obtain a cutout from the JWST/NIRCam imaging mosaic containing only the source and the surrounding area. The cutout is centered on the right ascension and declination of the source, as reported in the JADES photometric catalogs. The dimensions of the cutout are determined by the bounding box for the source, also reported in the JADES photometric catalogs. Due to requirements within \texttt{pysersic}, the cutout must be square; as such, the cutout dimensions are $m \times m$ pixels, where $m$ is the largest edge length reported by the bounding box coordinates. We set a minimum cutout size of $m_\mathrm{min} = 50$ pixels ($1.5\arcsec$), such that objects where all bounding box edges are $< m_\mathrm{min}$ have a cutout size of $50 \times 50$ pixels. 

Cutouts are obtained for the imaging mosaic, uncertainty mosaic, and segmentation map. The segmentation map denotes which pixels in the imaging mosaic correspond to any individual source, encoded by their JADES source IDs. A more detailed explanation of the process for generating this segmentation map is provided in \citet{Robertson2026}.

\subsection{Selecting the PSF for each source}\label{subsec:psf_selection}

We provide a PSF to \texttt{pysersic} for the convolution of each source being modeled. We utilize the mPSFs described in \S\ref{subsec:psf}. Due to the various JWST/NIRCam programs which stack to form the JADES imaging mosaics in any given filter, sources may be imaged by multiple observational programs, each with varying exposure times and position angles. As such, we must carefully select the mPSF which we provide to \texttt{pysersic} to ensure that the PSF closely matches the observational conditions under which the source was imaged.

For each source, we generate a list of program IDs that contribute to the stacked image of that source by decoding the bithash map described by \citet{Johnson2026} and \citet{Robertson2026}. From this list of program IDs, we preferentially select the program with the longest exposure time, as it will typically dominate the flux contribution in the stacked mosaic; for many sources, this selected program is JWST Program 1180 (PI: D. Eisenstein). Then, the mPSF corresponding to this selected program is supplied to \texttt{pysersic}. This selection process allows us to account for both varying position angles between observational programs, thereby closely matching the PSF to the conditions of the detection image of any individual source, and for variations in the full width at half-maximum for each filter's PSF, as shown in \citet{Johnson2026}. 

As a requirement, the input PSF array given to \texttt{pysersic} must be less than or equal to the size of the input image; as such, we crop the provided mPSF to the same dimensions as the input image, and check to ensure that the mPSF still normalizes to $\sim1$.

\subsection{Profile selection and setting priors}\label{subsec:profile_selection}

We model each source with a single-component S\'ersic profile by setting the profile type to \texttt{`sersic'} in \texttt{pysersic}, sampling for 7 different structural parameters: centroid values $x_c$ and $y_c$, total flux, effective radius along the semi-major axis $r_{\text{eff}}$, S\'ersic index $n$, ellipticity $e$, and position angle (PA) $\theta$.

For this single-component S\'ersic profile, we set prior distributions for the model parameters using the \texttt{autoprior} function within \texttt{pysersic}. The prior distributions for the centroid position and total flux are Gaussian distributions, centered on initial measurements made with \texttt{photutils}. The effective radius $r_{\text{eff}}$ prior distribution is a truncated Gaussian distribution, centered on an initial measurement from \texttt{photutils} and constrained to a lower limit of 0.5 pixels. The S\'ersic index $n$, ellipticity $e$, and PA $\theta$ all have uniform prior distributions, bounded as $n = [0.65, 8]$, $e = [0, 0.9]$, and $\theta = [0, 2\pi]$, respectively. We do not fit the sky background, as the segmentation map will be supplied as a mask to exclude nearby pixels whose flux is likely background-dominated.

The structural parameters reported in the catalogs described in \S\ref{sec:catalog} result from this single-component S\'ersic profile fitting to the JADES sources.

Additional profile types can be modeled, as well. In \S\ref{subsec:bd_evo}, we describe additional work to perform bulge-disk decomposition on a subset of sources in GOODS-S. To do this, we set the profile type to \texttt{`sersic\_exp'}, where the S\'ersic index of the disk component is set to $n_{\text{disk}}=1$ and the S\'ersic index of the bulge component is allowed to vary freely. In this profile, there are 10 different structural parameters: centroid values $x_c$ and $y_c$, total flux, effective radius of the bulge $r_{\text{eff,bulge}}$, effective radius of the disk $r_{\text{eff,disk}}$, S\'ersic index of the bulge $n_{\text{bulge}}$, ellipticity of the bulge $e_{\text{bulge}}$, ellipticity of the disk $e_{\text{disk}}$, bulge-to-total flux ratio $B/T$, and PA $\theta$.

For the bulge-disk profile, we once again employ the \texttt{autoprior} function from \texttt{pysersic} to set the prior distributions on the structural parameters. The prior distributions for the centroid position, total flux, and PA are repeated from the single-component S\'ersic priors. The prior distribution for $r_{\text{eff,bulge}}$ is a truncated Gaussian distribution, centered on an initial measurement from \texttt{photutils} multiplied by 2/3, and constrained to a lower limit of 0.5 pixels. For $r_{\text{eff,disk}}$, the prior distribution assumes this radius to be larger than $r_{\text{eff,bulge}}$; as such, the prior is set as a truncated Gaussian distribution, centered on the same initial measurement from \texttt{photutils} but now multiplied by 3/2, and constrained to a lower limit of 0.5 pixels. The prior distribution for $n_{\text{bulge}}$ is set as a truncated Gaussian distribution, centered on $n=4$, which defines the classical bulge-component. It is bounded to have lower and upper limits of 0.65 and 8, respectively. The priors for $e_{\text{bulge}}$ and $e_{\text{disk}}$ are equivalent uniform priors, bounded as $e_{\text{bulge}}, e_{\text{disk}} = [0,0.9]$. Lastly, the bulge-to-total flux ratio is defined by a uniform prior, varying as $B/T = [0,1]$.

\subsection{Deploying Bayesian modeling}\label{subsec:mcmc}

We utilize the Markov chain Monte Carlo (MCMC) sampling functionality of \texttt{pysersic}, which implements a No U-turn Sampler (NUTS; \citealt{nuts}) to sample the structural parameters of the morphological profile. 

The image, uncertainty, and segmentation map cutouts, along with the selected PSF, are provided to \texttt{pysersic}. The segmentation map cutout is supplied as a mask, such that only pixels which are associated with the source are considered in the \texttt{pysersic} modeling.

Using a Gaussian loss function, we perform MCMC sampling with 1 chain of 1,000 warm-ups and 250 samples. We select the median values from the posterior distributions of each parameter as the best-fit values for the model. All posterior samples are saved and reported in the catalog; a more detailed description of the content of the catalog follows in \S\ref{sec:catalog}. An exploration of the effect of varying the number of chains, warm-ups, and samples can be found in Appendix \ref{app:sampling_consistency}; we choose these quantities to consistently reproduce our results while minimizing overall computational expense.

We build a best-fit model using the \texttt{HybridRenderer} function in \texttt{pysersic}, which renders the S\'ersic profile using a hybrid real-Fourier algorithm introduced in \citet{render}. This algorithm balances the need to oversample central pixels in the high-S\'ersic index case with computational efficiency by representing the profile as a series of Gaussians in Fourier space. In Appendix \ref{app:rendering_consistency}, we explore the variability of our results for different rendering schemes implemented in \texttt{pysersic}.

\subsection{Injection-recovery simulations}\label{subsec:simulations}

\begin{figure*}[t]
    \centering
    \includegraphics[width=\textwidth]{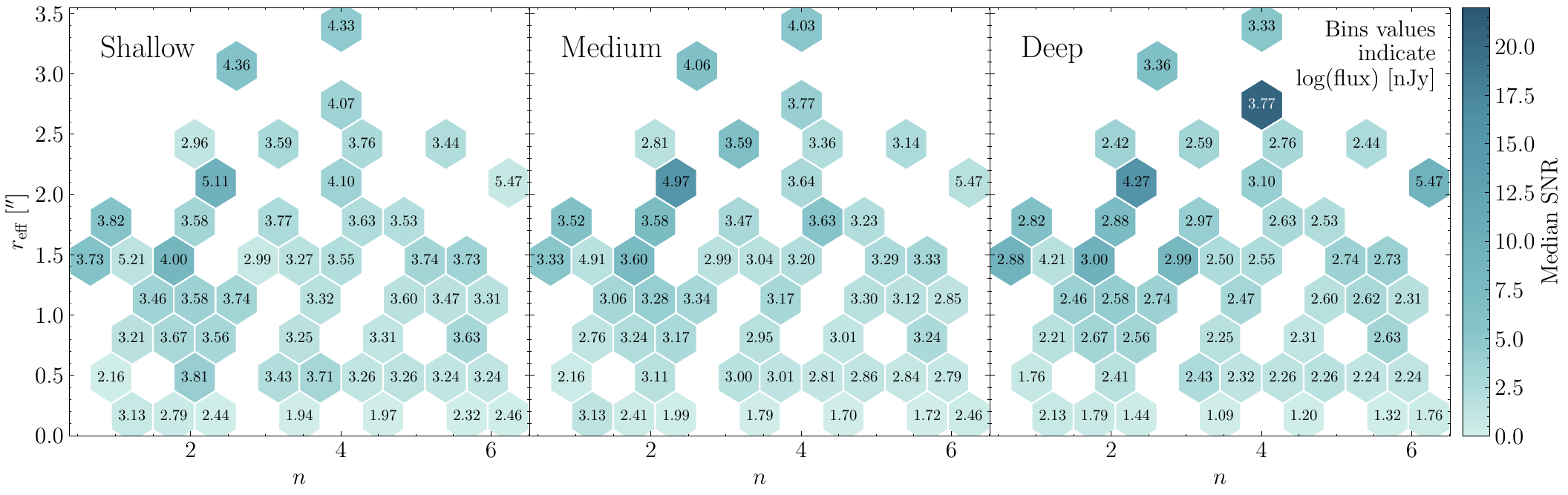}
    \caption{Injection-recovery simulation results from the (\textit{left}) shallow, (\textit{middle}) medium, and (\textit{right}) deep imaging regions, respectively, in the JWST/NIRCam F444W imaging mosaic in GOODS-S. Each bin value, in text, indicates the median log(flux) in nJy at which the flux value input into the model is sufficiently recovered by \texttt{pysersic}, for that $n$ and $r_{\text{eff}}$ bin. The color bar indicates the SNR of that recovered model, calculated as the total flux within the segmentation map of the input model, before being added to the noisy sky region, divided by the total noise flux of the sky region within the segmentation map. Darker blue colors indicate that a higher SNR is required to recover the correct flux of the source.}
    \label{fig:injection}
\end{figure*}

We perform injection-recovery simulations to understand the required flux and SNR thresholds to accurately recover the morphological parameters of each source across varying $n$ and $r_{\text{eff}}$ values. We focus this analysis on the JWST/NIRCam F444W imaging mosaic in GOODS-S.

We select a sample of real sources from the JADES DR5 photometric catalogs which, when modeled as described here, yield a representative distribution of values in $n$ and $r_{\text{eff}}$. We select 54 sources with $n=[0.5,6]$ and $r_{\text{eff}} = [0.2,3.5]\arcsec$. Because we will be generating a series of synthetic models from the fit parameters of these objects, we do not concern ourselves with the quality-of-fit, as compared to the JWST/NIRCam F444W imaging, for the selected sources. We perform the surface brightness profile modeling described in this section to return the median values of the posterior distributions for the structural parameters of these sources, and in so doing, generate the best-fit model describing them.

We then generate a series of synthetic models using the same median values of the structural parameters, but varying the total flux. We generate these models by multiplying the initial flux values by factors of [1, 2, 5, 7, 10, 20, 50, 70]. We convolve each model with the mPSF associated with the initial source, using the \texttt{convolve\_fft} function from \texttt{astropy} \citep{astropy_2013,astropy_2018}.

Each of these synthetic models are then added to empty regions in the JADES JWST/NIRCam F444W imaging mosaic in GOODS-S. We repeat this process three times for three separate empty regions, each representing different depths. Our shallow region has a median uncertainty value of 0.216 nJy, our medium depth region has a value of 0.097 nJy, and our deepest region has a value of 0.019 nJy.

We then refit all of these synthetic models, now added to each separate empty sky region, with the same modeling methodology described in this section. During this refitting, we use the segmentation map of the real source as a mask, such that the same number of pixels are utilized in the fitting of the synthetic models as in the initial model. Upon refitting, we analyze which models approximately return the initial structural parameter values with which the synthetic models were generated. We conclude that the recovery was successful if the range defined as $[flux_{\text{2.5\textsuperscript{th}percentile}}, flux_{\text{97.5\textsuperscript{th}percentile}}]$ for the initial model, scaled to the corresponding multiplicative factor, overlaps with the equivalent range for the refit model.

We report the minimum flux and SNR for which this recovery is achieved. The SNR is computed as the total flux within the segmentation map of the input model, without noise, divided by the total noise flux of the sky region within the segmentation map.

\begin{table*}[t]
    \centering
    \begin{tabular}{ccccc}
        \hline\hline
        Field & Filter& $N$ Sources in & $N$ Sources Detected & Percentage of Detected \\
        {} & {} & Morphology Catalogs & in Photometry & Sources Represented \\
        \hline
        GOODS-N & F070W & 34 931 & 37 176 & 94.0\% \\
        {} & F090W & 124 199 & 126 867 & 97.9\% \\
        {} & F115W & 146 185 & 160 485 & 91.1\% \\
        {} & F150W & 132 272 & 135 845 & 97.4\% \\
        {} & F200W & 137 506 & 154 943 & 88.7\% \\
        {} & F277W & 119 260 & 125 304 & 95.2\% \\
        {} & F356W & 136 828 & 147 265 & 92.9\% \\
        {} & F444W & 154 260 & 173 626 & 88.8\% \\
        \hline
        GOODS-S & F070W & 102 444 & 105 561 & 97.0\% \\
        {} & F090W & 222 042 & 226 290 & 98.1\% \\
        {} & F115W & 278 021 & 287 517 & 96.7\% \\
        {} & F150W & 284 470 & 291 324 & 97.6\% \\
        {} & F200W & 284 709 & 292 759 & 97.3\% \\
        {} & F277W & 282 952 & 290 282 & 97.5\% \\
        {} & F356W & 284 717 & 294 277 & 96.8\% \\
        {} & F444W & 294 248 & 297 335 & 99.0\% \\
        \hline
    \end{tabular}
    \caption{Number of objects, by field and by filter, reported in the morphological catalogs. The number of sources detected in the photometric catalogs \citep{Robertson2026}, computed as the number of objects for which the photometric \texttt{CIRC0} flux in that filter is non-zero, is also included.}
    \label{tab:catalog_fractions}
\end{table*}

In Figure \ref{fig:injection}, we show the results of these injection-recovery simulations. For the shallow, medium, and deep imaging regions, respectively, we show the median log(flux) in nJy (in text) and median SNR (in the color bar) for which the recovery was successful, binned by the $n$ and $r_{\text{eff}}$ values for those models. We find that, across all imaging regions, the flux of sources with higher S\'ersic indices are easier to recover than their low-$n$ counterparts. In particular, the combination of large-effective radius ($r_{\text{eff}}>1\arcsec$) and low-S\'ersic index ($n<3$) is challenging to recover, requiring total flux values $>10^3$ nJy for \texttt{pysersic} to accurately recover the total flux value in the deep imaging region. Conversely, compact sources with small-effective radius ($r_{\text{eff}}<0.5\arcsec$) and high-S\'ersic index ($n>3$) are well-recovered; even in the shallow imaging region, the flux of these compact sources can be recovered at values as low as 87 nJy.

These results indicate that the morphological fitting presented in this work may fail to accurately compute the total flux for large, extended, faint sources. Such sources have low levels of flux, tapering off exponentially, at large radii away from the peak of the surface brightness profile. These low-surface brightness outskirts are necessarily difficult to disentangle from sky background, presenting a physical limitation for morphological modeling.

\section{Catalogs of morphological properties in JADES}\label{sec:catalog}

We present catalogs of the morphological properties of $>10^5$ galaxies identified in the JWST/NIRCam wide-band imaging in both GOODS-N and GOODS-S, fitting single-component S\'ersic profiles to these sources. Table \ref{tab:catalog_fractions} reports the number of objects included in each catalog, by field and by filter. Each source is fit with the methodology described in \S\ref{sec:modeling}. Sources from the JADES photometric catalogs in these fields that are not included in the morphological catalogs suffered from failure modes which are explained in \S\ref{subsec:failures}.

The catalogs contain the best-fit (i.e., median of the posterior distribution) values for all morphological parameters, as well as the full posterior distributions which result from the MCMC sampling procedure. With these posterior distributions, we are able to calculate robust uncertainties for each morphological property, thereby improving upon other common methods for analyzing galaxy morphology. Additionally, the catalogs include all mPSFs used for convolution while fitting of the objects in the catalog as FITS ImageHDU extensions. All of these components of the catalogs are described in Appendix \ref{app:catalog_description}, Table \ref{tab:hdu}.

The best-fit models derived from the sampling and associated quality-of-fit statistics are reported in the \texttt{BEST\_FIT} HDU, described in Appendix \ref{app:catalog_description}, Table \ref{tab:bestfit}. As an example, in Figure \ref{fig:example}, we show the F444W imaging, best-fit S\'ersic model, and resulting residual for JADES ID 172089.

We perform unit conversions and reparameterizations on four of the morphological parameters to report in the final catalog. We convert the total flux from the JWST/NIRCam imaging units of MJy sr\textsuperscript{-1} to nJy via the \texttt{PIXAR\_SR} keyword in the header of the imaging. We convert the effective radius $r_{\text{eff}}$ from pixels to arcseconds by multiplying by 0.03$\arcsec$ pixel\textsuperscript{-1}, the pixel scale of the JADES imaging mosaics \citep{Johnson2026}. We reparameterize the ellipticity $e$ to the axis ratio $q$, which are related by $q = 1 - e$. Lastly, we convert the PA $\theta$ from radians to degrees.

\subsection{Failure modes}\label{subsec:failures}

\begin{figure*}[t]
    \centering
    \includegraphics[width=\textwidth]{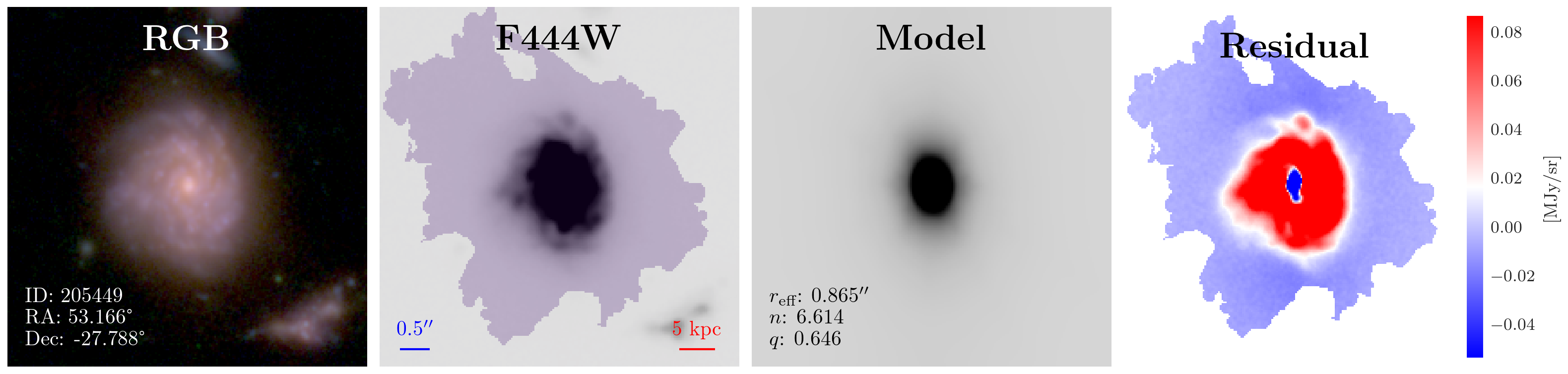}
    \caption{Example of an unsuccessful single-component S\'ersic fit to a spiral galaxy in GOODS-S. \textit{From left to right:} (1, left) F444W/F200W/F090W false-color RGB image of JADES ID 205449, with \zphot\ = 1.1 \citep{Robertson2026}. (2) F444W image of this object. Scale bars along the bottom indicate the size in angular and physical units, in blue and red, respectively. The purple shaded region represents the segmentation associated with this object. (3) The best-fit S\'ersic model for this object, as sampled by \texttt{pysersic}. Though JADES ID 205449 appears visually as a spiral galaxy, which commonly have S\'ersic indices $n\approx1$, the best-fit model finds $n=6.614$. (4, right) The residual of subtracting the S\'ersic model from the F444W image. Here, we find a case where a single-component S\'ersic model is clearly inadequate to describe this galaxy, given its bright nucleus alongside a flatter light distribution in the outer regions.}
    \label{fig:bad_example}
\end{figure*}

When attempting the modeling procedure described in \S\ref{sec:modeling}, for some sources, the sampling will fail to converge on a solution and result in a fit failure. These sources are therefore excluded from the morphological catalogs presented here. For this reason, the percentages representing the number of sources included in the morphological catalogs as a fraction of the total number of sources in the photometric catalogs are all $<100\%$.

The most common failure mode arises for sources which fall across or near the edge of the detector. In these cases, the surface brightness profile modeling cannot fit the radial profile in all directions and therefore does not produce a solution. To a lesser degree, sources falling near, but not across, a detector edge may not result in a sampling solution. In many cases, this failure seems to result from enhanced and non-uniform noise in the uncertainty image near detector edges.

\subsection{Caveats}\label{subsec:caveats}

In these catalogs, we report all sources which were successfully fit with a surface brightness profile and therefore did not suffer from any failure modes. However, we caution that there are sources in the catalog whose modeling did not fail, but whose returned model parameters may not be an accurate representation of the source. In these cases, the inferred structural parameters may not be astrophysically meaningful. An example of one such source can be seen in Figure \ref{fig:bad_example}; in this case, JADES ID 205449, we find that a single-component S\'ersic profile was successfully fit, but upon viewing the residual image, it is clear that the best-fit model resulting from the fitting does not capture the true surface brightness profile of the source. Given the high S\'ersic index ($n = 6.614$) of the best-fit model for this source, it is likely that a multiple-component fit, combining a point source or secondary S\'ersic component with a primary S\'ersic component, would more accurately model this galaxy.

To identify sources for which successful but inaccurate models were produced, we provide quality-of-fit statistics, described in \S\ref{subsubsec:quality}. In addition to using these statistics, other selection criteria can be employed to identify sources with representative best-fit models; for example, selecting only sources whose morphological total flux is in good agreement with the total Kron flux calculated photometrically may appropriately exclude the poorly-fit models of concern. 

\subsubsection{Quality-of-fit}\label{subsubsec:quality}

We report the total $\chi^2$ and \textit{p}-value to evaluate the quality-of-fit between the best-fit S\'ersic profile and the source image. We calculate the total $\chi^2$ values comparing the input image and best-fit model, weighted by the uncertainty image for the source. The \textit{p}-value associated with this $\chi^2$-value is calculated using the \texttt{scipy} \citep{scipy} implementation of a $\chi^2$ survival function, such that \textit{p}-values approaching 1 are indicative of models that are close representations of their input images.

\subsubsection{Limitations of the S\'ersic profile}\label{subsubsec:limitations}

We note that the total $\chi^2$ values and \textit{p}-values reported here indicate how precisely the source can be described by a single-component S\'ersic profile, which assumes the surface brightness of the galaxy varies smoothly with radius. However, many galaxies have non-smooth features, such as strong spiral arms, bars, or clumpy substructure, which cannot be precisely modeled by a S\'ersic profile. That being the case, the total $\chi^2$ values and \textit{p}-values reported can be misleading; galaxies with clear, non-smooth substructure are reported as having \textit{p}-values approaching 0 throughout these catalogs. For many of these cases, the best-fit S\'ersic profile reported here does accurately reflect the morphology of the galaxy, up to the limitations imposed by the assumptions of the S\'ersic profile. However, that accuracy may be underestimated if the total $\chi^2$ value and \textit{p}-value alone are used to make that assertion.

\subsection{Forthcoming catalogs of multi-component profiles}\label{subsec:forthcoming_catalogs}

Future work will include the release of additional JADES morphological catalogs, which will instead focus on the modeling of multiple-component surface brightness profiles. We aim to provide the structural parameters describing subsets of the JADES DR5 sources with bulge-disk decomposition, as well as joint S\'ersic-point source models, in multiple wide-band filters. We briefly explore performing bulge-disk decomposition on GOODS-S galaxies in \S\ref{subsec:bd_evo}.

\section{Rest-frame optical redshift evolution of morphology}\label{sec:redshift_evolution}

\begin{figure*}[t]
    \centering
    \includegraphics[width=\linewidth]{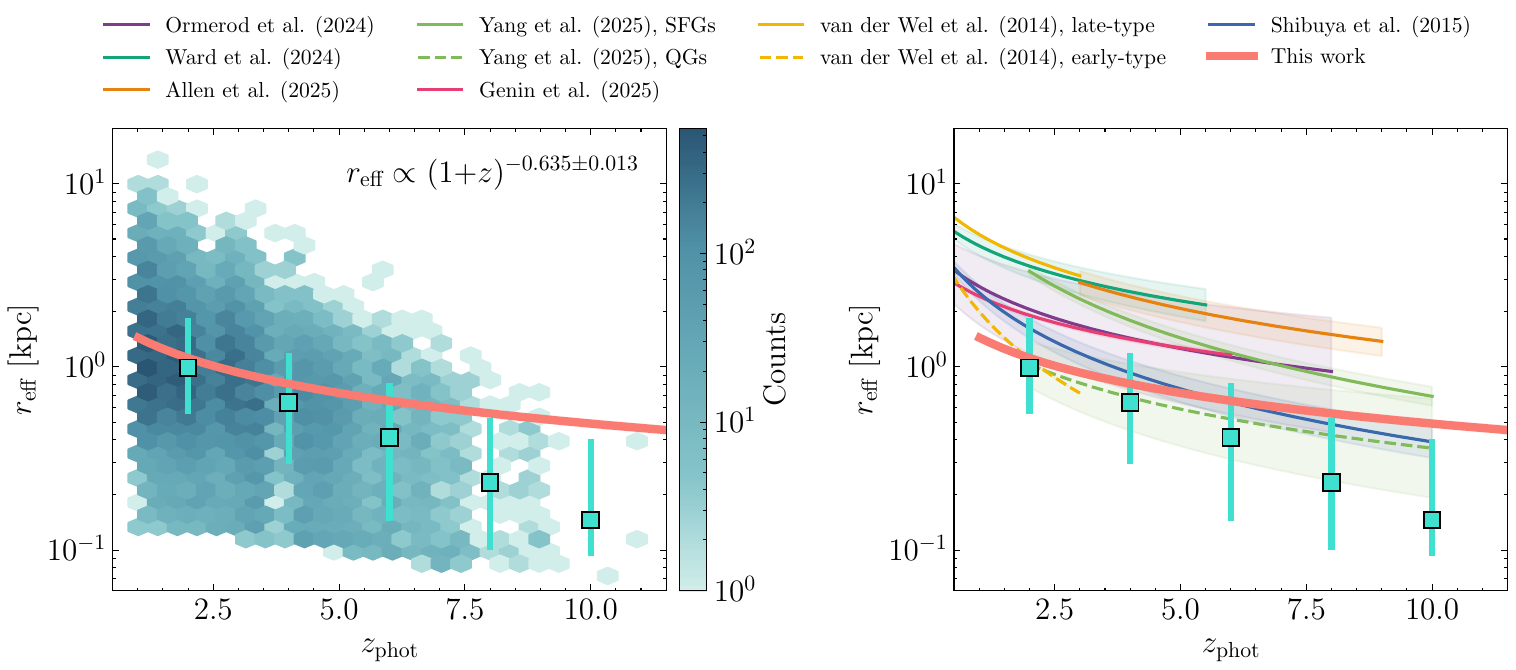}
    \caption{\textit{Left:} The rest-frame optical size evolution of the galaxies in our sample. We plot the data to highlight the density of galaxies, with darker blue colors representing a higher number of galaxies. The pink line shows the power-law parameterization fit to all data points, where $R=2.248 \pm 0.037$ kpc and $\beta_z= -0.635 \pm 0.013$. The turquoise points indicate the median $r_{\text{eff}}$ values in redshift bins of width $z=2$, with the error bars showing the 16\textsuperscript{th} and 84\textsuperscript{th} percentiles of the $r_{\text{eff}}$ distribution for that bin. \textit{Right:} The pink line, showing the power-law fit to our data, and the turquoise points, showing our sample binned by redshift, are repeated from the left panel. We plot the power-law parameterizations found in other JWST- and HST-era analyses of the size evolution of galaxies \citep{vdw_2014, shibuya_2015, ormerod_2024, ward_2024, allen_2025, yang_2025, genin_2025}; we note that the methodologies of each of these studies differ, which we discuss further in \S\ref{subsec:size_evo_discussion}.}
    \label{fig:r_eff_redshift_evo}
\end{figure*}

With these catalogs, we investigate the redshift evolution of the structural parameters for galaxies identified in JADES DR5 photometry. Throughout this work, we limit our analysis to the redshift evolution of the rest-frame optical ($\lambda = 5,000 \, \text{\AA}$). We convert angular sizes to physical sizes in kiloparsecs by using the photometric redshifts reported in \citet{Robertson2026}.

\subsection{S\'ersic evolution at $z > 1$}\label{subsec:sersic_evo}

Here, we analyze the rest-frame optical redshift evolution of the effective radius and luminosity surface density within a radius of 1 kiloparsec, derived from our single-component S\'ersic profiles.

We begin with an initial sample of 399,188 sources across GOODS-N and GOODS-S for which a rest-frame optical S\'ersic profile has been computed. We select only galaxies which are isolated, as their morphological fits are unlikely to be adversely affected by flux contamination from nearby sources. A galaxy is identified as isolated when it does not have any bright neighbors, reported as \texttt{FLAG\_BN=0}, and when it is not otherwise flagged in the rest-frame optical filter, as reported in the photometric catalogs by \citet{Robertson2026}. This selection removes 39.5\% of sources from the initial sample.

We select only sources which have well-constrained photometric redshifts spanning $1 < z_{\text{phot}} < 15$. We define the reported \zphot\ as well-constrained if the difference between the 16\textsuperscript{th} and 84\textsuperscript{th} percentile values of the posterior distribution for the photometric redshift \zphot\ is no more than 0.5. These criteria remove 72.2\% of the remaining sources. We exclude 276 other sources which were identified as stars, diffraction spike segments, brown dwarfs (as reported in \citealt{hainline_2024_bd, hainline_2025}), or other artifacts through a combination of excessive flux measurements and visual inspection.

We perform two cuts to remove sources whose S\'ersic profiles seem non-physical. First, we remove sources where the standard deviation of the posterior distribution for any structural parameter is 0; that is to say, sources where the MCMC sampling did not fully explore the prior distribution are removed. Second, to remove sources whose surface brightness profile fit does not accurately represent the source, as described in \S\ref{subsec:caveats}, we retain only the galaxies whose total Kron flux values reported by \citet{Robertson2026} agree with the total flux derived from the morphological fitting to within 10\%. These cuts remove 62.8\% of the remaining sources, leaving a sample of 24,905 sources. Finally, we remove any galaxies for which $r_{\text{eff}}<0.016\arcsec$, approximately half of the pixel scale of the JADES imaging mosaics, as this removes sources whose diameters are smaller than a pixel and therefore are functionally unresolved. This final cut removes 213 sources, amounting to 0.86\% of the remaining sources.

These selections result in a final subsample of 24,692 galaxies, across both GOODS-N and GOODS-S.

In Figure \ref{fig:r_eff_redshift_evo}, we show the rest-frame optical size evolution of our sample. These measurements indicate that the rest-frame optical effective radius decreases with increasing redshift; at $z\sim1$, we find $r_{\text{eff}}\sim 1.45$ kpc, increasing from $r_{\text{eff}}\sim 0.6$ kpc at $z\sim7$. We fit this size-redshift evolution with a power-law parameterization defined as
\begin{equation}
    r_{\text{eff}} = R (1 + z)^{\beta_z} \, \text{[kpc]}, 
\end{equation}
and find $R=2.248 \pm 0.037$ kpc and $\beta_z= -0.635 \pm 0.013$ for our sample. We note that the power-law fit does not go through the redshift-binned median $r_{\text{eff}}$ values shown in Figure \ref{fig:r_eff_redshift_evo}, especially at high-redshift; this occurs because there are significantly more sources at low-redshift than at high-redshift, so those sources dominate the power-law fit. In this sample, 45\% of the sources have \zphot\ $<2$, while $<1\%$ have \zphot\ $>7$.  In Figure \ref{fig:r_eff_redshift_evo}, we draw comparisons to the results of other HST and JWST surveys, and discuss further in \S\ref{subsec:size_evo_discussion} below.

In Figure \ref{fig:sigma_redshift_evo}, we also analyze the redshift evolution of the luminosity surface density within 1 kiloparsec, $\Sigma_{\text{1 kpc}}$. We find very little evolution with time; across redshift bins, the median values are all consistent to $\sim$1.5 MJy sr\textsuperscript{-1}. This implies that the core luminosity of galaxies increases in proportion to their sizes, thereby maintaining a relatively constant surface density across cosmic time. We note that cosmological surface brightness dimming affects this redshift evolution; we discuss this topic in \S\ref{subsec:size_evo_discussion}.

\begin{figure}
    \centering
    \includegraphics[width=\linewidth]{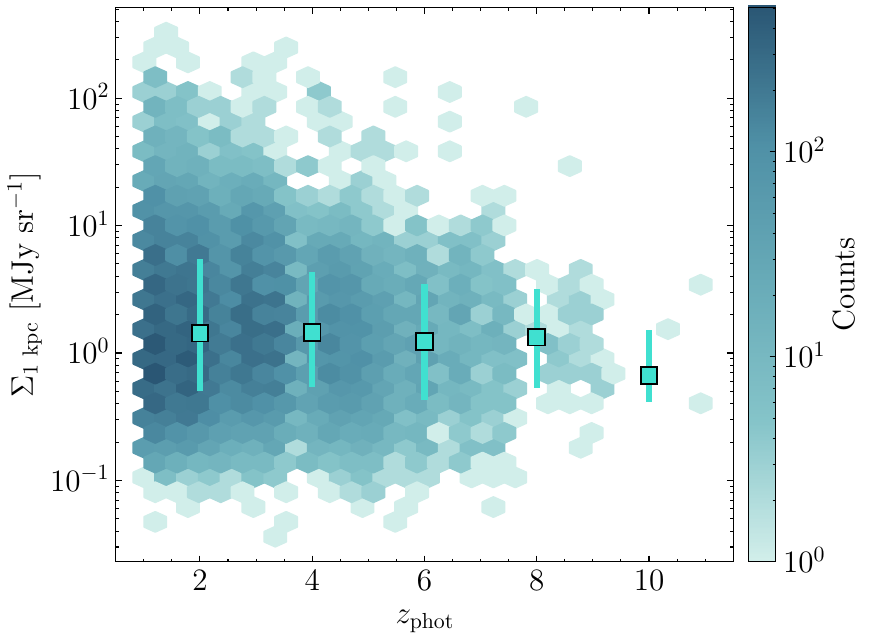}
    \caption{The luminosity surface density within 1 kiloparsec, $\Sigma_{\text{1 kpc}}$, as a function of redshift. We plot the data to represent the density of galaxies, with darker blue representing a higher number of galaxies. The turquoise points indicate the median $\Sigma_{\text{1 kpc}}$ values in redshift bins of width $z=2$, with the error bars showing the 16\textsuperscript{th} and 84\textsuperscript{th} percentiles of the $\Sigma_{\text{1 kpc}}$ distribution for that bin.}
    \label{fig:sigma_redshift_evo}
\end{figure}

\subsection{Bulge-disk decomposition evolution in JADES deep imaging}\label{subsec:bd_evo}

\begin{figure*}[t]
    \centering
    \includegraphics[width=\linewidth]{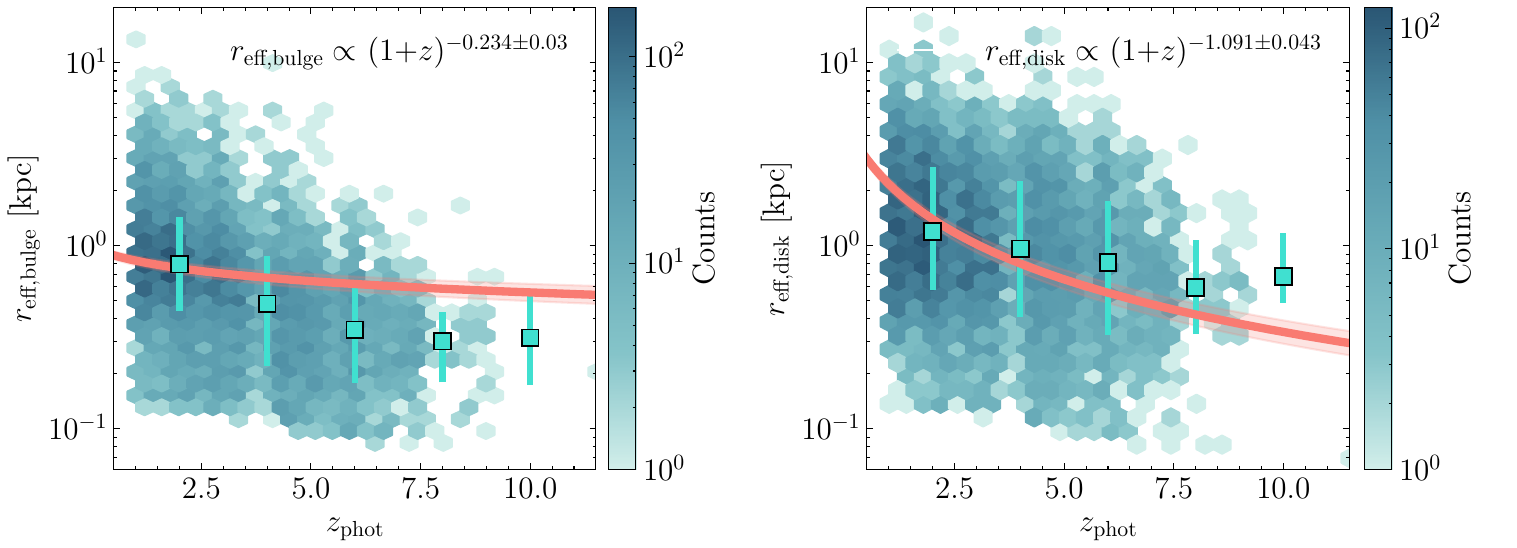}
    \caption{\textit{Left:} The rest-frame optical size evolution of the bulge-component for galaxies in our HUDF sample. We plot the data to represent the density of galaxies, with darker blue colors representing a higher number of galaxies. The pink line shows the power-law parameterization fit to all data points, where $R=0.975 \pm 0.036$ kpc and $\beta_z= -0.234 \pm 0.03$. The turquoise points indicate the median $r_{\text{eff,bulge}}$ values in redshift bins, with the error bars showing the 16\textsuperscript{th} and 84\textsuperscript{th} percentiles of the $r_{\text{eff,bulge}}$ distribution for that bin. \textit{Right:} The rest-frame optical size evolution of the disk-component for galaxies in our HUDF sample. The pink line shows the power-law parameterization fit to all data points, where $R=4.63 \pm 0.205$ kpc and $\beta_z= -1.091 \pm 0.043$. The turquoise points indicate the median $r_{\text{eff,disk}}$ values in redshift bins, with the error bars showing the 16\textsuperscript{th} and 84\textsuperscript{th} percentiles of the $r_{\text{eff,disk}}$ distribution for that bin.}
    \label{fig:bd_r_eff_redshift_evo}
\end{figure*}

We perform bulge-disk decomposition using the profile selection described in \S\ref{subsec:profile_selection}. We focus on $1<z<15$ sources which were observed by JWST Program 1180 (PI: D. Eisenstein), a region of deep imaging in GOODS-S overlapping the Hubble Ultra Deep Field (HUDF; \citealt{hudf_2006}). We limit ourselves to this region to leave an expanded analysis as future work, aiming instead to demonstrate the value of performing this detailed morphological fitting.

We begin with a sample of 15,611 galaxies in this region for which the total $\chi^2$ value for the bulge-disk profile is lower than the single-component S\'ersic profile in F444W, indicating that the multi-component fit is the better representation of the source. We note that we do not fix the S\'ersic index of the bulge-component to $n_{\text{bulge}}=4$, as is commonly done. The disk component is fixed to $n_{\text{disk}}=1$. As such, the analysis presented here does not assert that all sources in our sample are well-described by classical bulge- and disk-components; rather, we assert that these sources are more accurately represented by this multi-component fit than with a single-component S\'ersic profile. We use the terms ``bulge" and ``disk" to describe the inner and outer components of these multi-component profiles, respectively.

We once again identify galaxies which are isolated by including only sources with \texttt{FLAG\_BN=0}. As in the sample from \S\ref{subsec:sersic_evo}, we include only galaxies which are not otherwise flagged in their respective rest-frame optical filters, and remove sources where the standard deviation of the posterior distribution for any structural parameter is 0. We again require the photometric redshift to be well-constrained, as described in \S\ref{subsec:sersic_evo}, by requiring that $z_{\text{phot,84}}-z_{\text{phot,16}} < 0.5$. These cuts remove 6,857 sources, 43.9\% of our initial sample.

\begin{figure}
    \centering
    \includegraphics[width=\linewidth]{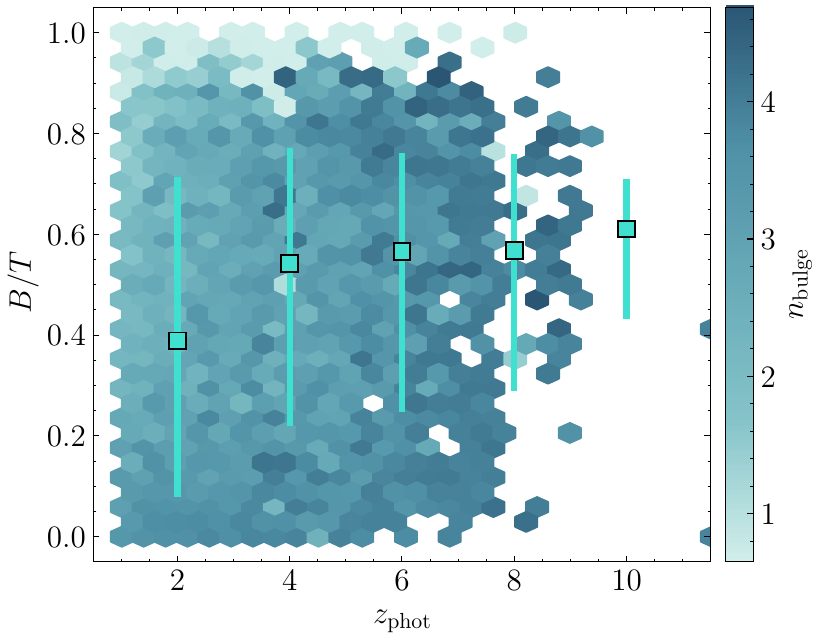}
    \caption{The flux ratio of the bulge relative to the total flux, $B/T$, as a function of redshift. The color bar indicates the S\'ersic index of the bulge component. The turquoise points indicate the median $B/T$ values in redshift bins of width $z=2$, with the error bars showing the 16\textsuperscript{th} and 84\textsuperscript{th} percentiles of the $B/T$ distribution for that bin.}
    \label{fig:bd_f_1_redshift_evo}
\end{figure}

We remove any galaxies for which the bulge or disk size is less than approximately half of the pixel scale of the JADES imaging mosaics, as they are functionally unresolved; that is to say, we remove sources with $r_{\text{eff,bulge}}<0.016\arcsec$ or $r_{\text{eff,disk}}<0.016\arcsec$. This removes 183 sources, which is 2.1\% of the sources which remained following the selections described above.

Lastly, we remove 181 sources identified as galaxies with clumpy substructure, rather than well-described S\'ersic components, from \citet{Zhu2026JADESClumps}. We note that our results are not impacted by the inclusion or exclusion of these sources. 

We conclude with a sample of 8,390 galaxies, for which we perform bulge-disk decomposition modeling in the rest-frame optical.

In Figure \ref{fig:bd_r_eff_redshift_evo}, we show the redshift evolution of the bulge- and disk-component effective radii, plotting also the power-law parameterization for each component. We find the bulge effective radii to evolve very little over time, evolving from $r_{\text{eff,bulge}} \sim 0.6$ kpc to $\sim 0.83$ kpc as the redshift goes from $z\sim 7$ to $z\sim 1$. The disk effective radii, in contrast, evolve rapidly. At $z\sim 7$, $r_{\text{eff,disk}}\sim 0.48$ kpc; by $z\sim 1$, $r_{\text{eff,disk}}$ has increased by a factor $4.5\times$ to $\sim 2.17$ kpc. In terms of the power-law parameters, we find that the size-evolution of the bulge-component effective radius can be described by
\begin{equation}
    r_{\text{eff,bulge}} = (0.975 \pm 0.036)(1+z)^{(-0.234 \pm 0.03)} \, \text{kpc},
\end{equation}
and the evolution of the disk-component effective radius can be described by
\begin{equation}
    r_{\text{eff,disk}} = (4.63 \pm 0.205)(1+z)^{(-1.091 \pm 0.043)} \, \text{kpc}.
\end{equation}

Furthermore, we analyze the ratio of the bulge-to-total flux, $B/T$. We trace the evolution of $B/T$ with redshift in Figure \ref{fig:bd_f_1_redshift_evo}, with the color bar indicating the S\'ersic index of the bulge component. We find that 34\% of our sample have $B/T > 0.6$, which is to say, they may be bulge-dominated. However, with this figure, we note that some degeneracies in the bulge-disk profile modeling emerge. For example, sources with $B/T \approx 1$ and $n_{\text{bulge}} \approx 1$ have essentially been modeled as single-component S\'ersic profiles with $n\approx1$. As such, the true percentage of bulge-dominated galaxies in this sample is $<34\%$. The median $B/T$ values, binned by redshift, do increase with increasing redshift; however, within their 16\textsuperscript{th} and 84\textsuperscript{th} percentiles, these binned values are all consistent with having no redshift evolution and therefore no particular epoch where bulge-dominated sources become dominant.

\section{Discussion}\label{sec:discussion}

In drawing comparisons to other works, we note the wide variety of morphological methodologies, selection criteria, and rest-frame wavelengths which have been utilized when analyzing the size evolution of galaxies. This variety may result in systematic differences in our results, thereby limiting the opportunity for direct comparison. That said, we attempt to draw some comparisons to other work and highlight relevant differences in approach, when needed.

Most notably, here, we do not select based on stellar mass. Given the scaling relationship between galaxy sizes and their stellar masses (e.g., \citealt{vdw_2014, shibuya_2015, ito_2024, ormerod_2024, varadaraj_2024, martorano_2024, ward_2024, morishita_2024, miller_2025, yang_2025, allen_2025, danhaive_2025}), we expect that our interpretation of our results would benefit from detailed analysis of the stellar mass completeness of our sample. Given the depth of the JADES imaging, we may be sensitive to lower mass sources than other studies; however, without analyzing the stellar masses of our sample, we cannot be certain. We also do not segregate our sample into star-forming versus quiescent galaxies, which may have varying effect at different redshifts. The number density of massive quiescent galaxies, for example, is significant at $z>1$, but has been found to decrease significantly by $z\sim5$ \citep{valentino_2023, baker_2025_quiescent, stevenson_2026}. As such, we expect that our sample of galaxies includes quiescent galaxies at low-redshift, which likely impacts our size evolution results, as we expect star-forming and quiescent galaxy sizes to evolve differently \citep{vdw_2014, ito_2024, ji_2024, baker_2025_sizemass, kawinwanichakij_2025, yang_2025}.

\subsection{Evaluating galaxy size evolution across surveys}\label{subsec:size_evo_discussion}

Several studies report the size evolution of galaxies by computing the intercept of the size-mass relation at a given mass. While this technique differs from the analysis performed here, we can compare values for the power-law index describing the size evolution, $\beta_z$. We found our value of $\beta_z= -0.635 \pm 0.013$ is within the range of values found in other studies. Two studies, \citet{ward_2024} and \citet{allen_2025}, employed multi-band fitting to calculate the effective radii of their sources. This approach jointly models the morphology in different bands and connects the results via a polynomial fit for each structural parameter. \citet{ward_2024} studies the rest-frame optical ($\lambda = 5,000 \, \text{\AA}$) size evolution via the intercept of the size-mass relation at $M_* = 5 \times 10^{10} M_\odot$, finding the intercept to be $R = 7.1 \pm 0.5$ kpc. The redshift range studied is somewhat smaller (out to $z \sim 5.5$), but nonetheless, they find $\beta_z = -0.63 \pm 0.07$, closely matching our results here. For a wider redshift range, \citet{allen_2025} reports the size evolution for a sample of star-forming galaxies at $3 \lesssim z \lesssim 9$, again derived from the intercept of the size-mass relation at $M_* = 5 \times 10^{10} M_\odot$. With this, they find $\beta_z = -0.807 \pm 0.026$ and $ R = 8.851 \pm 1.033$ kpc. These studies are broadly consistent with HST observations of the size evolution of late-type galaxies \citep{vdw_2014}.

Some studies have devised other methodologies for characterizing the effective radii, and therefore size evolution, of their data. \citet{genin_2025} finds $r_{\text{eff}} \propto (1+z)^{-0.59 \pm 0.01}$ kpc for a sample of log$(M_*/M_\odot) > 10$ galaxies, fitting their power-law parameterization to modeled S\'ersic profile $r_{\text{eff}}$ values, instead of calculating the intercept of the size-mass relation. Notably, \citet{genin_2025} reports S\'ersic profile sizes as the average size across imaging in the $\sim 1-5 \mu$m range, not in a single band, as has been done here. \citet{varadaraj_2024} analyzed the size evolution of massive, bright galaxies with log$(M_*/M_\odot) > 9$. They draw comparisons between different methodologies for binning down the sizes by redshift, using the peaks in the log-normal distribution in redshift bins, as introduced by \citet{curtis-lake_2016}, versus averaging across bins. Without binning their data at all, they find $\beta_z= -0.60 \pm 0.22$ in the rest-frame optical, with a higher normalization factor, $R=3.51$ kpc, than we find here. Despite utilizing different methodologies, \citet{genin_2025} and \citet{varadaraj_2024} show remarkable agreement with each other.

Similarly to the work performed here, \citet{ormerod_2024} fits S\'ersic profiles to massive (log$(M_*/M_\odot) > 9.5$) galaxies, fitting only the band corresponding to the rest-frame optical for each object. They calculate the parameters of the power-law as $ R = 4.50 \pm 1.32$ kpc and $\beta_z = -0.71 \pm 0.19$. Their value of $R$ is twice the value found in this study, though their $\beta_z$ values agrees with the value found in this study, within our uncertainties; this could be explained due to their mass selection, which is the only property of their study for which we have no point of comparison in our study.

With regard to the redshift evolution of $\Sigma_{\text{1 kpc}}$, we note that cosmological surface dimming causes the observed surface brightness to be reduced by a factor of $(1+z)^4$ \citep{tolman_1930, tolman_1934}. As such, we might expect $\Sigma_{\text{1 kpc}}$ to decrease with increasing redshift, maintaining this relationship. However, here we find that $\Sigma_{\text{1 kpc}}$ is relatively constant with increasing redshift. This may indicate that $\Sigma_{\text{1 kpc}}$ is intrinsically increased at higher redshift, and observational dimming effects cause the observed flat evolution. This would suggest that the core luminosity surface density of high-redshift galaxies is intrinsically higher than their low-redshift counterparts, possibly an indication of more star-formation activity in these regions \citep{robertson_2023}. However, additional analysis would be required to support this claim.

\subsection{Bulge-disk decomposition at high-redshift}\label{subsec:bd_evo_discussion}

Few studies have performed a detailed analysis of the redshift evolution of bulge-disk decomposition modeling at the redshifts analyzed here. \citet{genin_2025} performed both single-S\'ersic and bulge-disk decomposition on their sample. They find, similar to our results, that the disk component grows in radius rapidly, increasing as $\sim1.2$ kpc to $\sim3.4$ kpc from $z \sim 5$ to $z \sim 0.5$. Additionally, they find, in  agreement with our results, that the bulge-component effective radius is relatively constant, of typical size $<1$ kpc across time. They quantitatively classify their sample as being disk- or bulge-dominated, by drawing a comparison to the S\'ersic index of each source from a single-component fit alongside the calculated $B/T$ value. With this information, they compare the size evolution for each component for disk- and bulge-dominated galaxies separately, and assert that the disk-component of bulge-dominated galaxies may grow at later times. 

\section{Conclusions}\label{sec:conclusions}

This work presents catalogs of morphological properties for the sources detected in the JADES DR5 imaging mosaics across two legacy extragalactic fields. The deep, uniform imaging data from JADES is ideal for accurately modeling the surface brightness profiles of galaxies. We use \texttt{pysersic} to perform MCMC modeling of single-component S\'ersic profiles on all sources detected in the JADES DR5 photometry \citep{Robertson2026}, individually analyzing each galaxy in imaging mosaics from \citet{Johnson2026} for each of the eight JWST/NIRCam wide-band filters. This results in over 3 million individual S\'ersic models which we catalog here. In so doing, we have constructed one of the largest morphological datasets in the JWST-era, allowing for analysis of the individual rest-frame properties of each galaxy.

Beyond the sheer volume of sources modeled, we also have employed Bayesian inference to ensure the statistical quality of the information presented. These catalogs report the posterior distributions for each structural parameter for every S\'ersic profile computed. With this information, we aim to provide the community with a morphological dataset that enables detailed and statistically meaningful analysis across diverse galaxy populations.

To demonstrate the scientific opportunity made available by this dataset, we investigate the morphological evolution of $z>1$ galaxies in the rest-frame optical. We summarize our results as follows:

\begin{enumerate}
    \item In a sample of 24,692 galaxies, the rest-optical sizes of galaxies at high-redshift are typically smaller than those at low-redshift, with $r_{\text{eff}}\sim 0.6$ kpc at $z\sim7$ increasing to $r_{\text{eff}}\sim 1.45$ kpc by $z\sim1$, in good agreement with other studies.
    \item This evolution can be parameterized as a power-law, $r_{\text{eff}} \propto (1+z) ^{\beta_z}$. We find the power-law index to be $\beta_z= -0.635 \pm 0.013$, which is in reasonable agreement with other works. The normalization of the power-law  is found to be $R=2.248 \pm 0.037$ kpc, lower than the results of most other JWST-era analyses. However, we do not separate star-forming and quiescent galaxies from our sample. Therefore, direct comparison of the power-law normalization may be limited.
    \item For 8,390 galaxies in and surrounding HUDF, we perform bulge-disk decomposition to analyze the size evolution of multi-component surface brightness profiles. We find that bulge-component sizes are relatively constant across time, increasing $\sim 0.23$ kpc from $z \sim 7$ to $\sim 1$. Meanwhile, the disk-component radius grows rapidly, increasing by a factor of $4.5\times$ in that same period.
\end{enumerate}

In the future, we aim to combine this morphological information with catalogs of other galaxy properties to broadly explore the evolution of galaxies over cosmic time. We plan to release catalogs of the structural parameters for multi-component surface brightness profile modeling, which we believe will enable more granular analysis of the morphological evolution of high-redshift galaxies. These datasets, when taken together, may provide answers to our outstanding questions on the nature of galaxy formation in the early universe, and the evolution of galaxies from early times through today.

\begin{acknowledgments}

This work is based on observations made with the NASA/ESA/CSA James Webb Space Telescope. The data were obtained from the Mikulski Archive for Space Telescopes at the Space Telescope Science Institute, which is operated by the Association of Universities for Research in Astronomy, Inc., under NASA contract NAS 5-03127 for JWST.

JADES DR5 includes JWST/NIRCam data from JWST Programs 1176, 1180, 1181, 1210, 1264, 1283, 1286, 1287, 1895, 1963, 2079, 2198, 2514, 2516, 2674, 3215, 3577, 3990, 4540, 4762, 5398, 5997, 6434, 6511, and 6541.

The authors acknowledge the teams of Programs 1895, 1963, 2079, 2514, 3215, 3577, 3990, 6434, and 6541 for developing their observing program with a zero-exclusive-access period.

The authors acknowledge use of the lux supercomputer at UC Santa Cruz, funded by NSF MRI grant AST 1828315.

CC and BER acknowledge support from the JWST/NIRCam Science Team contract to the University of Arizona, NAS5-02105, and JWST Programs 3215 and 5015. DJE, ZJ, BDJ, MR, CNAW, and YZ also acknowledge support from the JWST/NIRCam Science Team contract to the University of Arizona, NAS5-02105. AJB and JC acknowledge funding from the ``FirstGalaxies" Advanced Grant from the European Research Council (ERC) under the European Union’s Horizon 2020 research and innovation programme (Grant agreement No. 789056). AJC gratefully acknowledges support from the Cosmic Dawn Center through the DAWN Fellowship. The Cosmic Dawn Center (DAWN) is funded by the Danish National Research Foundation under grant No. 140. The research of CCW is supported by NOIRLab, which is managed by the Association of Universities for Research in Astronomy (AURA) under a cooperative agreement with the National Science Foundation. ST acknowledges support by the Royal Society Research Grant G125142. Funding for the research of RH was provided by the Johns Hopkins University, Institute for Data Intensive Engineering and Science (IDIES). WMB gratefully acknowledges support from DARK via the DARK fellowship. This work was supported by a research grant (VIL54489) from VILLUM FONDEN. ECL acknowledges support of an STFC Webb Fellowship (ST/W001438/1). ALD thanks the University of Cambridge Harding Distinguished Postgraduate Scholars Programme and Technology Facilities Council (STFC) Center for Doctoral Training (CDT) in Data intensive science at the University of Cambridge (STFC grant number 2742605) for a PhD studentship. H\"U acknowledges funding by the European Union (ERC APEX, 101164796). Views and opinions expressed are however those of the authors only and do not necessarily reflect those of the European Union or the European Research Council Executive Agency. Neither the European Union nor the granting authority can be held responsible for them. RM acknowledges support by the Science and Technology Facilities Council (STFC), by the ERC through Advanced Grant 695671 “QUENCH”, and by the UKRI Frontier Research grant RISEandFALL. RM also acknowledges funding from a research professorship from the Royal Society. DJE is supported as a Simons Investigator. JAAT acknowledges support from the Simons Foundation and JWST Program 3215. AU acknowledges support by the National Science Foundation Graduate Research Fellowship Program under Grant No. 2240310. Any opinions, findings, and conclusions or recommendations expressed in this material are those of the author(s) and do not necessarily reflect the views of the National Science Foundation.

Support for JWST Program 3215 was provided by NASA through a grant from the Space Telescope Science Institute, which is operated by the Association of Universities for Research in Astronomy, Inc., under NASA contract NAS 5-03127.

\end{acknowledgments}

\begin{contribution}

CC performed the surface brightness profile fitting, catalog construction, and morphological analysis  described in this work, in addition to writing the text. BDJ and BER performed the imaging reduction and photometric analysis, respectively, on which the morphological analysis was performed.


\end{contribution}

\facilities{JWST(NIRCam)}

\software{astropy \citep{astropy_2013,astropy_2018}, pysersic \citep{pysersic}, jax \citep{jax2018github}, numpyro \citep{numpyro_2021, numpyro_2025}, photutils \citep{photutils}}

\appendix
\restartappendixnumbering

\section{Detailed catalog description}\label{app:catalog_description}

In the tables below, we describe the content and format of the morphological catalogs presented here. In particular, we describe the HDUs which comprise the catalogs in Table \ref{tab:hdu}, and we list the columns, units, and descriptions of the \texttt{BEST\_FIT} HDU in Table \ref{tab:bestfit}.

\begin{table*}[t!]
    \centering
    \begin{tabular}{cccl}
        \hline\hline
        Ext. Number & Name & Type & Description \\
        \hline
        0 & \texttt{PRIMARY} & PrimaryHDU & Header contains relevant package version numbers and specifies the \\
        {} & {} & {} & JADES photometric catalog on which this catalog is based. \\
        1 & \texttt{BEST\_FIT} & BinTableHDU & Contains the median values from the posterior distributions of each \\
        {} & {} & {} & morphological parameter and associated quality-of-fit information. \\
        {} & {} & {} & See Table \ref{tab:bestfit} for a more detailed description. \\
        2 & \texttt{MEAN\_STD} & BinTableHDU & Contains the mean and standard deviation values from the \\
        {} & {} & {} & posterior distributions of each morphological parameter. \\
        3 & \texttt{PERCENTILES} & BinTableHDU & Contains the median and 5\textsuperscript{th}, 16\textsuperscript{th}, 84\textsuperscript{th}, and 95\textsuperscript{th} percentiles from \\
        {} & {} & {} & the posterior distributions of each morphological parameter. \\
        4 & \texttt{XC\_SAMPLES} & BinTableHDU & Contains the 250 samples comprising the \texttt{XC} posterior distribution, \\
        {} & {} & {} & in units of pixels. \\
        5 & \texttt{YC\_SAMPLES} & BinTableHDU & Contains the 250 samples comprising the \texttt{YC} posterior distribution, \\
        {} & {} & {} & in units of pixels. \\
        6 & \texttt{FLUX\_SAMPLES} & BinTableHDU & Contains the 250 samples comprising the \texttt{FLUX} posterior \\
        {} & {} & {} & distribution, in units of nJy. \\
        7 & \texttt{N\_SAMPLES} & BinTableHDU & Contains the 250 samples comprising the \texttt{N} posterior distribution, \\
        {} & {} & {} & which are unitless. \\
        8 & \texttt{R\_EFF\_SAMPLES} & BinTableHDU & Contains the 250 samples comprising the \texttt{R\_EFF} posterior \\
        {} & {} & {} & distribution, in units of arcseconds. \\
        9 & \texttt{Q\_SAMPLES} & BinTableHDU & Contains the 250 samples comprising the \texttt{Q} posterior distribution, \\
        {} & {} & {} & which are unitless. \\
        10 & \texttt{PA\_SAMPLES} & BinTableHDU & Contains the 250 samples comprising the \texttt{PA} posterior distribution, \\
        {} & {} & {} & in units of degrees, East of North. \\
        11 & \texttt{PSF\_LIST} & BinTableHDU & Contains an integer indicating which appended ImageHDU mPSF \\
        {} & {} & {} & was supplied to \texttt{pysersic} for image convolution. For example, an \\
        {} & {} & {} & object with a value of 12 in the  \texttt{PSF\_ENCODING} column utilized \\
        {} & {} & {} & the mPSF appended as \texttt{PSF\_12}. \\
        12... & \texttt{PSF\_12...} & ImageHDU & Appended as separate ImageHDUs, we include all mPSFs that \\
        {} & {} & {} & were supplied to \texttt{pysersic} for the fitting of sources in the catalog. \\
        {} & {} & {} & The integer subscript in the HDU Name corresponds to the \\
        {} & {} & {} & encoding integer reported in \texttt{PSF\_LIST}. \\
        \hline
    \end{tabular}
    \caption{List of all HDUs in each morphological catalog presented here, along with a description of the HDU contents.}
    \label{tab:hdu}
\end{table*}

\begin{table*}[t!]
    \centering
    \begin{tabular}{ccl}
        \hline\hline
        Column Name & Units & Description \\
        \hline
        \texttt{ID} & N/A & JADES DR5 source ID. \\
        \texttt{PID} & N/A & JADES DR5 parent ID, i.e., the parent object from which the source is deblended. \\
        \texttt{RA} & degree & Right ascension of the source, reported in ICRS J2000. \\
        \texttt{DEC} & degree & Declination of the source, reported in ICRS J2000. \\
        \texttt{X} & pixel & Pixel coordinate from the JADES DR5 imaging mosaic corresponding to the listed \\
        {} & {} & RA/Dec of the source. \\
        \texttt{Y} & pixel & Pixel coordinate from the JADES DR5 imaging mosaic corresponding to the listed \\
        {} & {} & RA/Dec of the source. \\
        \texttt{XC} & pixel & Median of the posterior distribution for the \texttt{X} pixel coordinate of the light centroid of \\
        {} & {} & the source, as reported by \texttt{pysersic}. \\
        \texttt{YC} & pixel & Median of the posterior distribution for the \texttt{Y} pixel coordinate of the light centroid of \\
        {} & {} & the source, as reported by \texttt{pysersic}. \\
        \texttt{FLUX} & nJy & Median of the posterior distribution for the total flux of the source, as reported by \\
        {} & {} &  \texttt{pysersic}, and converted from units of MJy sr\textsuperscript{-1} to nJy. \\
        \texttt{N} & N/A, unitless & Median of the posterior distribution for the S\'ersic index $n$ of the source, as reported \\
        {} & {} &  by \texttt{pysersic}. \\
        \texttt{R\_EFF} & arcsecond & Median of the posterior distribution for the effective radius $r_{\text{eff}}$ of the source, as \\
        {} & {} & reported by \texttt{pysersic}, and converted from units of pixels to arcseconds. \\
        \texttt{Q} & N/A, unitless & Median of the posterior distribution for the axis ratio of the source, reparameterized \\
        {} & {} &  from the ellipticity reported by \texttt{pysersic}. \\
        \texttt{PA} & degree &  Median of the posterior distribution for the position angle of the source, as reported \\
        {} & {} & by \texttt{pysersic}, and converted from units of radians to degrees.\\
        \texttt{CUTOUT\_DIM0} & pixel & The 0\textsuperscript{th}-dimension size of the cutout array supplied to \texttt{pysersic} for the source. \\
        \texttt{CUTOUT\_DIM1} & pixel & The 1\textsuperscript{st}-dimension size of the cutout array supplied to \texttt{pysersic} for the source. \\
        \texttt{NPIX} & N/A & Number of pixels in the source cutout which are considered during the fitting. This \\
        {} & {} & value is less than \texttt{CUTOUT\_DIM0} $\times$ \texttt{CUTOUT\_DIM1} due to masking, as described \\
        {} & {} & in \S\ref{subsec:mcmc}. \\
        \texttt{CHI\_TOTAL} & N/A & Total $\chi^2$ value comparing the original image cutout to the rendered model utilizing \\
        {} & {} & the parameter values listed in this table, as described in \S\ref{subsubsec:quality}. \\
        \texttt{P\_VALUE} & N/A & \textit{p}-value associated with the reported total $\chi^2$ value, as described in \S\ref{subsubsec:quality}.\\
        \hline
    \end{tabular}
    \caption{List of all columns in \texttt{BEST\_FIT} HDU, along with a description of the column contents.}
    \label{tab:bestfit}
\end{table*}

\section{MCMC sampling consistency}\label{app:sampling_consistency}

Here, we demonstrate the consistency of the resulting structural parameters from the S\'ersic profile fitting when varying the number of warmups, samples, and chains in the MCMC sampling mode from \texttt{pysersic}. For this analysis, we utilize the same subsample of galaxies as described in \S\ref{subsec:sersic_evo} and compare the structural parameters as modeled on the JADES DR5 JWST/NIRCam F444W imaging mosaic from \citet{Johnson2026}.

In \texttt{pysersic}, the MCMC sampling mode defaults to sampling with 2,000 warmups and 2,000 samples divided between two chains. In this work, we reduce those quantities to 1,000 warmups and 250 samples on only one chain. The reduced sampling minimizes computational expense, and yields consistent results, as we show here.

\begin{figure}
    \centering
    \includegraphics[width=\linewidth]{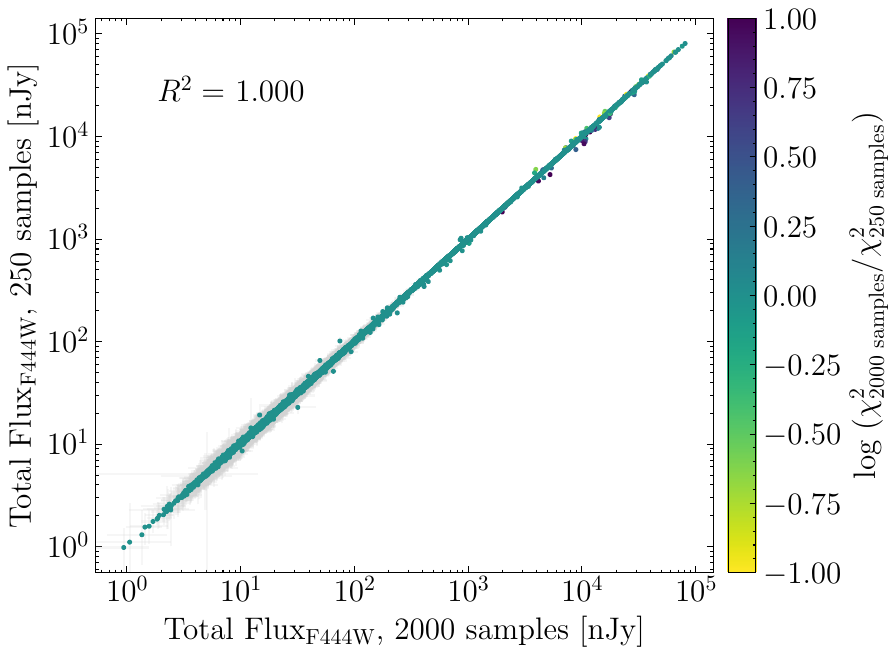}
    \caption{Comparison of the total flux computed by \texttt{pysersic} for the default number of samples (2,000) versus the number of samples computed in this work (250). The gray error bars show the 16\textsuperscript{th} and 84\textsuperscript{th} percentiles for each parameter value, and the points are colored according to the ratio of the total $\chi^2$ values from each model.}
    \label{fig:appendix_samp_flux}
\end{figure}

\begin{figure}
    \centering
    \includegraphics[width=\linewidth]{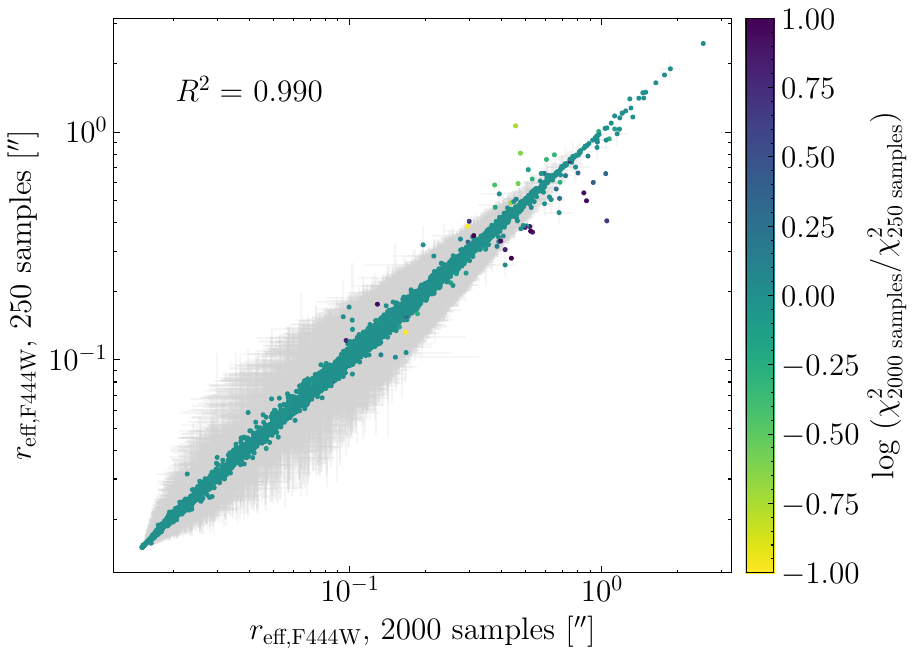}
    \caption{Comparison of the effective radius computed by \texttt{pysersic} for the default number of samples (2,000) versus the number of samples computed in this work (250). The error bars and color bar are the same as described in Figure \ref{fig:appendix_samp_flux}.}
    \label{fig:appendix_samp_reff}
\end{figure}

\begin{figure}
    \centering
    \includegraphics[width=\linewidth]{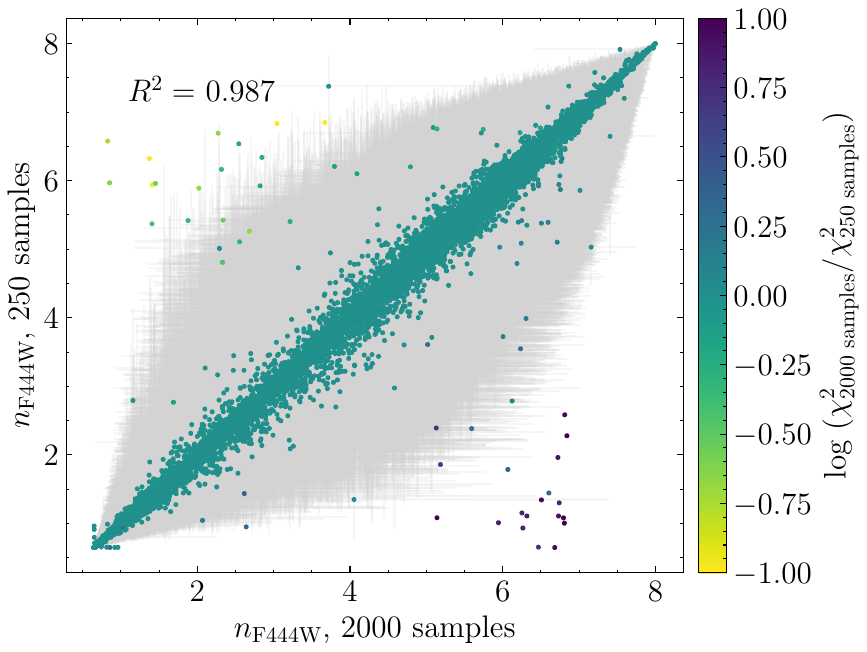}
    \caption{Comparison of the S\'ersic index $n$ computed by \texttt{pysersic} for the default number of samples (2,000) versus the number of samples computed in this work (250). The error bars and color bar are the same as described in Figure \ref{fig:appendix_samp_flux}.}
    \label{fig:appendix_samp_n}
\end{figure}

Figures \ref{fig:appendix_samp_flux}, \ref{fig:appendix_samp_reff}, and \ref{fig:appendix_samp_n} show the results obtained for the total flux, effective radius, and S\'ersic index, respectively, for this sample of galaxies. The gray error bars indicate the 16\textsuperscript{th} and 84\textsuperscript{th} percentiles for each parameter value. The color bar shows the ratio of the total $\chi^2$ value for the model with 2,000 samples to the total $\chi^2$ value for 250 samples, in log scale; color bar values $>0$ indicate that the model derived from 2,000 samples has a higher total $\chi^2$ value, and is therefore a worse representation of the imaged source, than the total $\chi^2$ for the model with 250 samples.

In general, we find the resulting structural parameters to be consistent between sampling methods, with $R^2 =$ 1.000, 0.990, and 0.987 for the flux, $r_{\text{eff}}$, and $n$ values, respectively. For $r_{\text{eff}}$ and $n$, where we find mild scatter off of the main relation, the scattered points do not indicate an overall preference between the models with respect to their $\chi^2$ values. Additionally, the uncertainties returned from each methodology are consistent; that it to say, obtaining more samples does not result in significantly improved uncertainty constraints, even though the posterior distribution is more densely populated in that case.

\section{Model rendering consistency}\label{app:rendering_consistency}

We demonstrate the consistency of the resulting structural parameters from the S\'ersic profile fitting for different rendering implementations available in \texttt{pysersic}. We once again utilize the same subsample of galaxies as described in \S\ref{subsec:sersic_evo}, comparing the structural parameters as modeled on the JADES DR5 JWST/NIRCam F444W imaging mosaic from \citet{Johnson2026}.

In \texttt{pysersic}, the default rendering algorithm implemented is called the \texttt{HybridRenderer}. This implementation, described earlier in \S\ref{subsec:mcmc}, renders the S\'ersic profile in a hybrid real-Fourier space which balances the need for computational efficiency with the accuracy of the resulting model. An alternative implementation, called with \texttt{PixelRenderer}, is a traditional algorithm where the profile is rendered in real space, with oversampling in the center to preserve accuracy in high-S\'ersic index cases. Each of these rendering algorithms are described in \citet{pysersic}. Using the \texttt{PixelRenderer} is significantly more computationally expensive; in some cases, particularly for larger galaxies, the sampling time using the \texttt{PixelRenderer} is over ten times longer than using the \texttt{HybridRenderer}, totaling over one hour to fit one source. Given the number of S\'ersic profiles we compute here, this additional computation expense would be prohibitive. To motivate the use of the \texttt{HybridRenderer} in this work, we compare the resulting morphological parameters for each rendering algorithm.

\begin{figure}
    \centering
    \includegraphics[width=\linewidth]{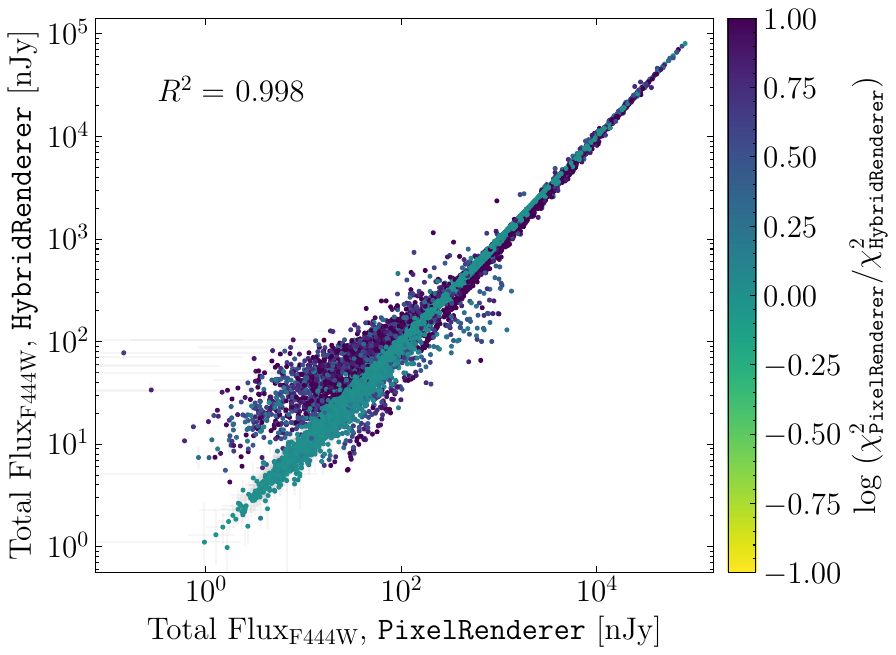}
    \caption{Comparison of the total flux computed by \texttt{pysersic} using the \texttt{PixelRenderer} versus the \texttt{HybridRenderer}. The gray error bars show the 16\textsuperscript{th} and 84\textsuperscript{th} percentiles for each parameter value, and the points are colored according to the ratio of the total $\chi^2$ values from each model.}
    \label{fig:appendix_rend_flux}
\end{figure}

\begin{figure}
    \centering
    \includegraphics[width=\linewidth]{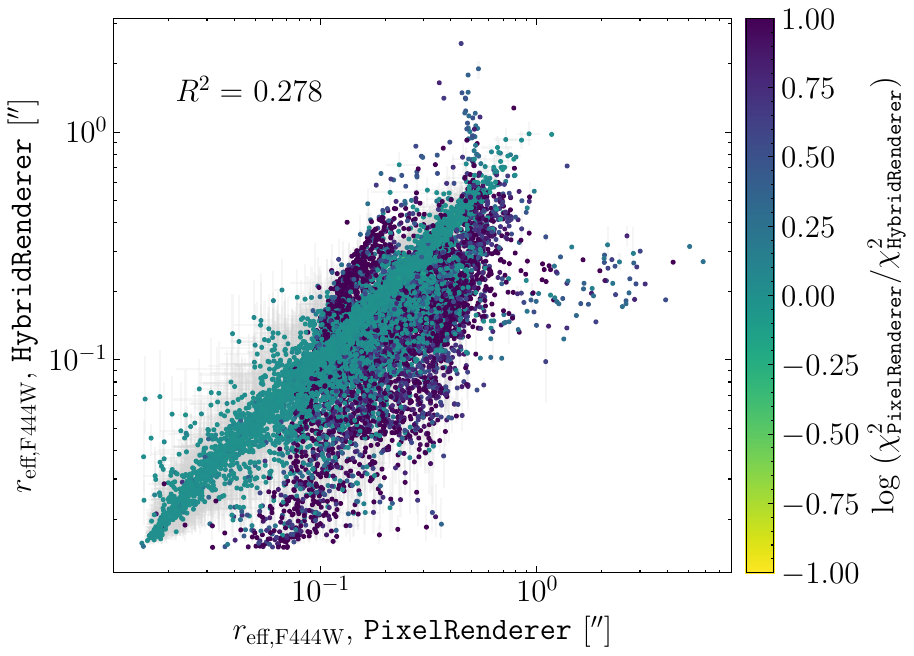}
    \caption{Comparison of the effective radius computed by \texttt{pysersic} using the \texttt{PixelRenderer} versus the \texttt{HybridRenderer}. The error bars and color bar are the same as described in Figure \ref{fig:appendix_rend_flux}.}
    \label{fig:appendix_rend_reff}
\end{figure}

\begin{figure}
    \centering
    \includegraphics[width=\linewidth]{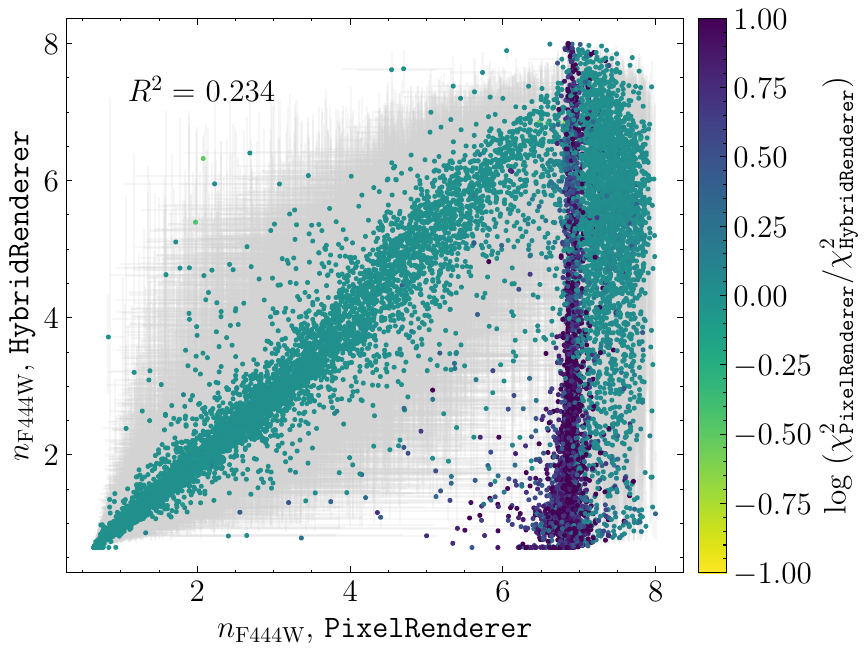}
    \caption{Comparison of the S\'ersic index $n$ computed by \texttt{pysersic} using the \texttt{PixelRenderer} versus the \texttt{HybridRenderer}. The error bars and color bar are the same as described in Figure \ref{fig:appendix_rend_flux}.}
    \label{fig:appendix_rend_n}
\end{figure}

Figures \ref{fig:appendix_rend_flux}, \ref{fig:appendix_rend_reff}, and \ref{fig:appendix_rend_n} show the results obtained for the total flux, effective radius, and S\'ersic index, respectively, for this sample of galaxies. The gray error bars indicate the 16\textsuperscript{th} and 84\textsuperscript{th} percentiles for each parameter value. The color bar shows the ratio of the total $\chi^2$ value for the model rendered by the \texttt{PixelRenderer} to the total $\chi^2$ value for the \texttt{HybridRenderer} model, in log scale; color bar values $>0$ indicate that the \texttt{PixelRenderer} has a higher total $\chi^2$ value, and is therefore a worse representation of the imaged source, than the total $\chi^2$ for the \texttt{HybridRenderer} model.

Unlike in Appendix \ref{app:sampling_consistency}, the resulting structural parameters are not well-correlated between rendering algorithms in every case. The total flux results appear to be consistent, with $R^2=0.998$. Interestingly, the objects which scatter away from the main relation are predominantly sources with log$(\chi^2_{\texttt{PixelRenderer}}/\chi^2_{\texttt{HybridRenderer}}) > 0$, indicating that these sources are more accurately modeled by the \texttt{HybridRenderer} model. The correlation for the effective radius and S\'ersic index values is poorer, with $R^2=0.278$ and $R^2=0.234$, respectively. In the case of the effective radius, we again find that objects which scatter off of the main relation seem to generally favor the \texttt{HybridRenderer} model as the more accurate model, though there is scatter for objects with no model preference, as well. For the S\'ersic index, 33.9\% of the sources find $n_{\texttt{PixelRenderer}}>6$, most of which do not agree with the computed $n_{\texttt{HybridRenderer}}$. Among those sources, 97.3\% have log$(\chi^2_{\texttt{PixelRenderer}}/\chi^2_{\texttt{HybridRenderer}}) > 0$, once again indicating that the \texttt{HybridRenderer} models are preferred in these cases where the parameter values otherwise do not agree.

Though the parameter values derived from each rendering algorithm are not always consistent with each other, in cases where they disagree, the $\chi^2$ values from each model indicate that the \texttt{HybridRenderer} rendering better captures the structural parameters of the source image. This motivates our use of the \texttt{HybridRenderer} algorithm throughout this work.

\bibliography{main}{}
\bibliographystyle{aasjournalv7}



\end{document}